%% file: main.tex
\documentclass[fleqn,usenatbib,useAMS]{mnras}
\usepackage{newtxtext,newtxmath}

\usepackage[T1]{fontenc}

\DeclareRobustCommand{\VAN}[3]{#2}
\let\VANthebibliography\thebibliography
\def\thebibliography{\DeclareRobustCommand{\VAN}[3]{##3}\VANthebibliography}

\usepackage{graphicx}	
\usepackage{amsmath}	

\usepackage{graphicx}
\usepackage[usenames,dvipsnames,table]{xcolor}
\usepackage{hyperref}
\usepackage{booktabs}
\usepackage{float}
\usepackage{multicol}
\usepackage{longtable}
\usepackage{threeparttablex}
\usepackage{caption}

\newcommand{\dex}{\mathrm{dex}}
\newcommand{\nm}{\mathrm{nm}}
\newcommand{\citoiv}{C\,\textsc{\lowercase{I}}\,--\,\textsc{\lowercase{IV}}}

\title{Extended theoretical transition data in \citoiv{}}

\author[W. Li et al.]{
W. Li,$^{1,3}$\thanks{E-mail: wenxian.li@mau.se}
A. M. Amarsi,$^{2}$
A. Papoulia$^{1,3}$
J. Ekman$^{1}$
and P. J\"onsson$^{1}$
\\
$^{1}$Department of Materials Science and Applied Mathematics, Malm\"o University, SE-205 06, Malm\"o, Sweden\\
$^{2}$Theoretical Astrophysics, Department of Physics and Astronomy, Uppsala University, Box 516, SE-751 20 Uppsala, Sweden\\
$^{3}$Division of Mathematical Physics, Lund University, Post Office Box 118, SE-221 00 Lund, Sweden
}

\date{Accepted XXX. Received YYY; in original form ZZZ}

\pubyear{2020}

\begin{document}
\label{firstpage}
\pagerange{\pageref{firstpage}--\pageref{lastpage}}
\maketitle

\begin{abstract}
Accurate atomic data are essential for opacity calculations and for abundance analyses of the Sun and other stars. The aim of this work is to provide accurate and extensive results of energy levels and transition data for \citoiv{}.

The Multiconfiguration Dirac–Hartree–Fock and relativistic configuration interaction methods were used in the present work. To improve the quality of the wave functions and reduce the relative differences between length and velocity forms for transition data involving high Rydberg states, alternative computational strategies were employed by imposing restrictions on the electron substitutions when constructing the orbital basis for each atom and ion.

Transition data, e.g., weighted oscillator strengths and transition probabilities, are given for radiative electric dipole (E1) transitions involving levels up to $\mathrm{1s^22s^22p6s}$ for \ion{C}{I}, up to $\mathrm{1s^22s^27f}$ for \ion{C}{II}, up to $\mathrm{1s^22s7f}$ for \ion{C}{III}, and up to $\mathrm{1s^28g}$ for \ion{C}{IV}. Using the difference between the transition rates in length and velocity gauges as an internal validation, the average uncertainties of all presented E1 transitions are estimated to be 8.05\%, 7.20\%, 1.77\%, and 0.28\%, respectively, for \citoiv{}. Extensive comparisons with available experimental and theoretical results are performed and good agreement is observed for most of the transitions. In addition, the \ion{C}{I} data were employed in a reanalysis of the solar carbon abundance.  The new transition data give a line-by-line dispersion similar to the one obtained when using transition data that are typically used in stellar spectroscopic applications today.

\end{abstract}

\begin{keywords}
Atomic data --- Atomic processes --- Line: formation ---  Radiative transfer --- Sun: abundances --- Methods: numerical
\end{keywords}



\section{Introduction}\label{sec:intro}

Accurate atomic data are of fundamental importance to
many different fields of astronomy and astrophysics.  
This is particularly true for carbon.  As the fourth-most
abundant metal in the cosmos \citep{2009ARA&amp;A..47..481A},
carbon is a major source of opacity in the atmospheres
and interiors of stars.
Complete and reliable sets of atomic data for carbon
are essential for stellar opacity calculations,
because of their significant impact on 
stellar structure and evolution
\citep[e.g.][]{2012ApJ...755...15V,2020ApJ...889..157C}.

Accurate atomic data for carbon are also important in the context 
of spectroscopic abundance analyses and 
Galactic Archaeology. Carbon abundances measured  
in late-type stars help us to understand the
nucleosynthesis of massive stars and AGB stars,
and thus the Galactic chemical evolution
\citep[e.g.][]{2020ApJ...888...55F,2020A&A...633L...9J,2020AJ....159...90S}.
In early-type stars, carbon abundances
help constrain the present-day Cosmic Abundance Standard
\citep[e.g.][]{2008A&A...481..199N,2012A&A...539A.143N,2019ApJ...884..150A}.
In the Sun, 
the carbon abundance is precisely measured 
in order to put different cosmic objects onto a common scale 
\citep[e.g.][]{2010A&A...514A..92C,
2019A&A...624A.111A}.
In all of these cases, oscillator strengths 
for \ion{C}{I} (cool stars) and for \citoiv{} (hot stars)
underpin the spectroscopic analyses; 
this is especially the case for studies
that relax the assumption of local thermodynamic equilibrium
\citep[LTE; e.g.][]{2001A&A...379..936P,2006ApJ...639L..39N},
in which case much larger sets of reliable atomic data are needed.

On the experimental side, a number of studies of transition data have been
presented in the literature.  Neutral \ion{C}{I} transition probabilities
for the 
$\mathrm{2p4p \rightarrow 2p3s}$ transition array have been studied by
\cite{Miller1974} using a spectroscopic shock tube and by \cite{Jones1984}
using a wall-stabilized arc. The measurements of relative oscillator strengths
for $\mathrm{2p3p \rightarrow 2p3s}$, $\mathrm{2p3d \rightarrow 2p3p}$ and
$\mathrm{2p4s \rightarrow 2p3p}$ have been performed by
\cite{MUSIELOK1997395,Bacawski_2001,Golly_2003} using a wall-stabilized arc.
Older measurements of oscillator strengths are also available
using the same technique
\citep{1953ZPhy..135...13M,1958ZPhy..151..114R,1962PPS....79...94F,1963ZNatA..18.1107B,1987A&A...181..203G,1989A&A...221..155G}.
By analysing the high-resolution spectra obtained
with the Goddard High Resolution Spectrograph on the Hubble Space Telescope,
\cite{Federman_2001} derived oscillator strengths for \ion{C}{I} lines below
1200 \AA.

For \ion{C}{II}, a number of measurements have also been performed. \cite{Trabert_1999}
measured the radiative decay rates for the intercombination (IC) transitions $\mathrm{2s2p^2~^4P \rightarrow 2s^22p~^2P^o}$ at a heavy-ion storage ring, and the total measured radiative decay rates to the ground term were 125.8 $\pm$ 0.9 s$^{-1}$ for $\mathrm{^4P_{1/2}}$, 9.61 $\pm$ 0.05 s$^{-1}$ for $\mathrm{^4P_{3/2}}$, and 45.35 $\pm$ 0.15 s$^{-1}$ for $\mathrm{^4P_{5/2}}$. The aforementioned results are, however, not in agreement with the values measured by \cite{Fang1993} using a radio-frequency ion trap, i.e., 146.4(+8.3, -9.2) s$^{-1}$ for $\mathrm{^4P_{1/2}}$, 11.6(+0.8, -1.7) s$^{-1}$ for $\mathrm{^4P_{3/2}}$, and 51.2(+2.6, -3.5) s$^{-1}$ for $\mathrm{^4P_{5/2}}$. 
\cite{Goly1982} measured the transition probabilities from a helium-carbon arc for some multiplets of $\mathrm{\{2p^3,2s^23p\} \rightarrow 2s2p^2}$ and $\mathrm{2s^24s \rightarrow 2s^23p}$ with estimated relative uncertainty of 50\%.
Using an electric shock tube, \cite{Roberts1967} provided the relative oscillator strengths of some \ion{C}{II} multiplets with relative uncertainties of 7\%.
\cite{Reistad_1986} gave lifetimes for 11 \ion{C}{II} levels using the beam-foil excitation technique and extensive cascade analyses.

For \ion{C}{III}, the IC decay rate of the $\mathrm{2s2p~^3P^o_1 \rightarrow 2s^2~^1S_0}$ transition was measured to be 121.0 $\pm$ 7 s$^{-1}$ by \cite{Kwong1993} using a radio-frequency ion trap and 102.94 $\pm$ 0.14 s$^{-1}$ by \cite{Doerfert1997} using a heavy-ion storage ring. 
The discrepancy between the values obtained from the two different methods is quite large, i.e., of the order of 15\%. The result given by the latter measurement is closer to earlier $ab~initio$ calculations ranging between 100 and 104~s$^{-1}$ \citep{Fleming_1994,Fischer_1994,Ynnerman1995}.
Several measurements have also been performed for the lifetimes of the
low-lying levels of \ion{C}{III}
\citep{Reistad1986,MICKEY197077,Nandi_1996,Buchet_Poulizac_1973}.

For the system of Li-like \ion{C}{IV}, the transition probabilities of the $\mathrm{1s^22p~^2P^o_{1/2,3/2} \rightarrow 1s^22s~^2S_{1/2}}$ transitions were 
measured by \cite{BERKNER196535} using the foil-excitation technique and by
\cite{KNYSTAUTAS197175} using the beam-foil technique, respectively.  There are
also a number of measurements of lifetimes in \ion{C}{IV}
using the beam-foil technique
\citep{Donnelly1978,Buchet1973,Jacques1980}.

On the theoretical side, Froese Fischer et al. 
have performed detailed studies of
\citoiv{},
focusing on the low-lying levels.
They carried out
Multiconfiguration Hartree-Fock (MCHF)
calculations and used the Breit-Pauli 
(MCHF-BP) approximation for computing energy levels and transition properties, e.g., transition probabilities, oscillator strengths, and lifetimes, in \ion{C}{I} \citep{Tachiev2001,Fischer_2006,FROESEFISCHER20041}, \ion{C}{II} \citep{Tachiev_2000}, \ion{C}{III} \citep{Tachiev_1999,Fischer_2000}, and \ion{C}{IV} \citep{Godefroid_2001,FISCHER1998119}. 

Hibbert et al. have presented extensive calculations 
for optical transitions.  They
used the CIV3 code \citep{HIBBERT1975141} to calculate
oscillator strengths and transition probabilities in \ion{C}{I}
\citep{Hibbert1993}, \ion{C}{II} \citep{CORREGE200419}, and \ion{C}{III}
\citep{Kingston_2000}. In the calculations of \cite{Hibbert1993,CORREGE200419},
empirical adjustments were introduced to the diagonal matrix elements in order
to accurately reproduce energy splittings.  
Their \ion{C}{I} oscillator strengths are frequently used
in the abundance analyses of cool stars
(Sect. \ref{solaranalysis}).

A number of other authors have also presented 
theoretical transition data for carbon.
\cite{Zatsarinny_2002} calculated
the oscillator strengths for transitions to high-lying excited states of
\ion{C}{I} using a spline frozen-cores method.  \cite{Nussbaumer1984} provided
the radiative transition probabilities using the $LS$-coupling approximation
and intermediate coupling approximation, respectively, for the six
energetically lowest configurations of \ion{C}{I}.  \cite{Nussbaumer1981}
calculated the transition probabilities for \ion{C}{II}, from terms up to
$\mathrm{2s^24f~^2F^o}$, using the $LS$-coupling and close coupling (CC)
approximation, respectively.

In view of the great astrophysical interest for
large sets of homogeneous atomic data,
extensive spectrum calculations of transition
data in the carbon atom and carbon ions were carried out 
under the umbrella of the Opacity Project 
using the CC approximation of the
R-matrix theory, and the results are available in the Opacity Project online
database (TOPbase; \cite{1992RMxAA..23..107C,1993A&A...275L...5C}).  The latest compilation of
\ion{C}{I} transition probabilities was made available by \cite{Haris_2017},
and those of \ion{C}{II-IV} can be found in earlier compilations by
\cite{Erratum,Wiese2007} and \cite{Fuhr2005}.

In this context,
the General-purpose Relativistic Atomic Structure Package (\textsc{Grasp}) 
has, more recently, been used by \cite{Aggarwal_2015} 
to predict the
radiative decay rates and lifetimes of 166 levels belonging to the n $\le$ 5
configurations in \ion{C}{III}. Using an updated and extended version of 
this code (\textsc{Grasp2K}),
\cite{JONSSON2010} determined transition data involving 26 levels in
\ion{C}{II}.

Although for the past decades a considerable amount of research has been
conducted for carbon, there is still a need for 
extended sets of reliable theoretical transition data.
To address this, we have carried out new calculations based on the
fully relativistic Multiconfiguration
Dirac–Hartree–Fock (MCDHF) and relativistic configuration interaction (RCI)
methods, as implemented in 
the newest version of the \textsc{Grasp} code, \textsc{Grasp2018} \citep{GraspV3,Grasp2018}.
We performed energy spectrum calculations for 100, 69, 114, and
53 states, in \citoiv{}, respectively.
Electric dipole (E1) transition data 
(wavelengths, transition probabilities, line strengths, 
and oscillator strengths) were computed along with the
corresponding lifetimes of these states.  

This paper is structured into six sections, including the introduction.  Our
theoretical methods are described in Sect. \ref{sec:theory}, and computational
details are given in Sect. \ref{sec:schemes}. In Sect. \ref{sec:results}, we
present our results and the validation of the data. 
As a complementary method of validation,
in Sect.
\ref{solaranalysis}, we use the derived data in a 
reanalysis of the solar carbon abundance.
Finally, we present our conclusions in Sect. \ref{sec:conclusion}.

\section{Theory}\label{sec:theory}
In the Multiconfiguration Dirac-Hartree-Fock (MCDHF) method \citep{Grant2007,Fischer2016}, wave functions for atomic states
$\gamma^{(j)}\, P JM$, $j=1,2,\ldots,N$
with angular momentum quantum numbers $JM$ and parity $P$ are expanded over {${N_\mathrm{CSFs}}$ configuration state functions}
\begin{equation}
    \Psi(\gamma^{(j)}\, P JM) = \sum_i^{{N_\mathrm{CSFs}}} c^{(j)}_i\, \Phi(\gamma_i\, P JM).
    \label{eq:csfs}
\end{equation}
The configuration state functions (CSFs) are $jj$-coupled many-electron functions, recursively built
from products of one-electron Dirac orbitals. As for the notation,
 $\gamma_i$ specifies the occupied subshells of the CSF with their complete angular coupling tree information.
The radial large and small components of the one-electron orbitals and the expansion coefficients \{$ c^{(j)}_i $\} of the CSFs 
are obtained, for a number of targeted states, by solving the Dirac-Hartree-Fock radial equations and the configuration interaction eigenvalue problem resulting from applying the variational principle on the statistically weighted energy functional of the targeted states with terms added for preserving the
orthonormality of the one-electron orbitals. 
The energy functional is based on the  Dirac-Coulomb (DC) Hamiltonian
and accounts for relativistic kinematic effects. 

Once the radial components of the one-electron orbitals are determined, higher-order interactions,
such as the transverse photon interaction and quantum electrodynamic (QED) effects (vacuum polarization
and self-energy), are added to the Dirac-Coulomb Hamiltonian.
Keeping the radial components fixed, the expansion coefficients  \{$ c^{(j)}_i $\} of the CSFs 
for the targeted states are obtained by solving the configuration interaction eigenvalue problem.

The evaluation of radiative E1 transition data (transition probabilities, oscillator
strengths) between two states: $\gamma' P'J'M'$ and $\gamma PJM$ is non-trivial. 
The transition data can be expressed in terms of reduced matrix elements of the transition operator ${\bf T}^{(1)}$:
\begin{eqnarray}
\langle \,\Psi(\gamma PJ)\, \|  {\bf T}^{(1)} \| \,\Psi(\gamma' P'J')\, \rangle  &=&
\nonumber \\
 \sum_{j,k} c_jc'_k \; \langle \,\Phi(\gamma_j PJ)\, \|  {\bf T}^{(1)} \| \,\Phi(\gamma'_k P'J')\, \rangle,
\end{eqnarray}
where $c_j$ and $c'_k$ are, respectively, the expansion coefficients of the CSFs for the lower and upper states, and the summation occurs over all the CSFs for the lower and upper states.
The reduced matrix elements are expressed via spin-angular coefficients $d^{(1)}_{ab}$ and operator strengths as:
\begin{eqnarray}
\langle \,\Phi(\gamma_j PJ)\, \|  {\bf T}^{(1)} \| \,\Phi(\gamma'_k P'J')\, \rangle &=&
\nonumber \\
 \sum_{a,b} d^{(1)}_{ab} \; \langle \,n_al_aj_a\, \|  {\bf T}^{(1)} \| \,n_bl_bj_b\, \rangle.
\end{eqnarray}
Allowing for the fact that we are now using Brink-and-Satchler type reduced matrix elements, we have
\begin{eqnarray}
\langle \,n_al_aj_a\, \|  {\bf T}^{(1)} \| \,n_bl_bj_b\, \rangle  &=&
\nonumber \\
\left( \frac{(2j_b + 1)\omega}{\pi c} \right) ^{1/2} (-1)^{j_a - 1/2}  \begin{pmatrix}j_a~~~~1~~~~j_b \\ \frac{1}{2}~~~~0~-\frac{1}{2}\end{pmatrix} \overline{M_{ab}}, 
\end{eqnarray}
where $\overline{M_{ab}}$ is the radiative transition integral defined by \cite{gauge}. The factor in front of $\overline{M_{ab}}$ is the Wigner 3-j symbol that gives the angular part of the matrix element.
The $\overline{M_{ab}}$ integral can be written
$\overline{M_{ab}} = \overline{M^e_{ab}} + G\overline{M^l_{ab}}$, where
$G$ is the gauge parameter. When $G=0$ we get the Coulomb gauge, whereas for $G=\sqrt{2}$ we get the Babushkin gauge.
The Babushkin gauge corresponds to the length gauge in the non-relativistic limit and puts weight on the outer part
of the wave functions \citep{gauge,Hibbert_1974}.
The Coulomb gauge corresponds to the velocity gauge and puts more weight on the inner part of the wave functions \citep{Papoulia2019}.
For E1 transitions, the Babushkin and Coulomb gauges give the same
value of the transition moment for exact solutions of the Dirac-equation \citep{gauge}.
For approximate solutions, the transition moments differ, and the quantity $dT$, defined as \citep{Fischer2009,Ekman2014}
\begin{equation}
\label{accuracy}
dT = \frac{|A_l-A_v|}{\max(A_l,A_v)},
\end{equation}
where $A_l$ and $A_v$ are transition rates in length and velocity form, can be
used as an estimation of the uncertainty of the computed rate.

\section{Computational schemes}\label{sec:schemes}
Calculations were performed in the extended optimal level (EOL)
scheme \citep{EOL} for the weighted average of the even and odd parity states.
The CSF expansions were determined using the multireference-single-double (MR-SD) method,
allowing single and double (SD) substitutions from 
a set of important configurations, referred to as the MR, to orbitals in an active set (AS) \citep{Olsen_AS,Sturesson_AS,Fischer2016}.
The orbitals in the AS 
are divided into spectroscopic orbitals, which build 
the configurations in the MR, and correlation orbitals, which are introduced to correct the initially obtained wave functions.
During the different steps of the calculations for \citoiv{}, the CSF expansions 
were systematically enlarged by adding layers of correlation orbitals.

MCDHF calculations aim to generate an orbital set. The orbital set is then used
in RCI calculations based on CSF expansions that can be enlarged to capture additional electron correlation effects. For the same CSF expansion,
different orbital sets give different results for both energy levels and
transition data. Conventionally, MCDHF calculations are performed for CSF
expansions obtained by allowing substitutions not only from the valence
subshells, but also from the subshells deeper in the core, accounting for
valence-valence (VV), core-valence (CV), and core-core (CC) electron
correlation effects. 
Using orbital sets from such calculations,
\cite{Pehlivan_MgI} predicted large $dT$ values for transitions between
low-lying states and high Rydberg states, indicating substantial uncertainties
in the corresponding transition data.  For transitions involving high Rydberg
states, it was shown that the velocity gauge gave the more accurate results,
which is contradictory to the general belief that the length gauge is the
preferred one \citep{Hibbert_1974}.  
Analyzing the situation more carefully,
\cite{Papoulia2019} found that correlation orbitals resulting from MCDHF
calculations based on CSF expansions obtained by allowing substitutions from
deeper subshells are very contracted in comparison with the outer Rydberg
orbitals. As a consequence, the outer parts of the wave functions for the
Rydberg states are not accurately described.  Thus, the length form that probes
the outer part of the wave functions does not produce trustworthy results,
while the velocity form that probes the inner part of the wave functions yields
more reliable transition rates. In the same work, the authors showed how transition rates that are only weakly sensitive to the choice of gauge can be obtained, by
paying close attention to the CSF generation strategies for the MCDHF
calculations.  

In the present work, following the suggestion by
\cite{Papoulia2019}, the MCDHF calculations were based on CSF expansions for
which we impose restrictions on the substitutions from the inner subshells and
obtain, as a consequence, correlation orbitals that overlap more with the
spectroscopic orbitals of the higher Rydberg states, adding to a better
representation of the outer parts of the corresponding wave functions. The MR
and orbital sets for each atom and ion are presented in Table~\ref{tab:MR}. The
computational scheme, including CSF generation strategies, for each atom and
ion is discussed in detail below.  The MCDHF calculations were followed by RCI
calculations, including the Breit interaction and leading QED effects. 

\input{scheme}

\subsection{\ion{C}{I}}
As seen in Table~\ref{tab:MR}, in the computations of neutral carbon, configurations with ${n = 7~(l = \mathrm{s}); 6~(l = \mathrm{p, d})}$, which are not of direct relevance, were included in the MR set to obtain orbitals that are spatially extended, improving the quality of the outer parts of
the wave functions of the higher Rydberg states. 
The MCDHF calculations were performed using CSF expansions that were produced by SD substitutions from the valence orbitals of the configurations
in the MR to the active set of orbitals,
with the restriction of allowing maximum one substitution from orbitals with $n=2$. The $\mathrm{1s^2}$ core was kept closed and, at this point, the expansions of the atomic states accounted for VV electron correlation. 
As a final step, an RCI calculation was performed for the largest SD valence
expansion augmented by a CV expansion.
The CV expansion was obtained by allowing SD substitutions
from the valence orbitals and the $\mathrm{1s^2}$ core of the configurations in the MR, with the restriction
that there should be at most one substitution from $\mathrm{1s^2}$.
The numbers of CSFs in the final even and odd state expansions are, respectively, 14 941 842 and 15 572 953, distributed over the different $J$ symmetries.

\subsection{\ion{C}{II}}
Similarly to the computations in \ion{C}{I}, in the computations of the singly-ionized carbon, the configurations $\mathrm{2s^2\{8s,8p,9s,9p\}}$, which are not our prime targets, were included in the MR set (see also Table~\ref{tab:MR}). In this manner, we generated orbitals that are localized farther from the atomic core.
The MCDHF calculations were performed using CSF expansions obtained by
allowing SD substitutions from the valence orbitals of the MR configurations. During this stage, the $\mathrm{1s^2}$ core remained frozen and the CSF expansions accounted for VV correlation.
The final wave functions of the targeted states were determined in an RCI calculation, which included CSF expansions that were formed by allowing SD substitution from all subshells of the MR
configurations, with the restriction that there should be at most
one substitution from the $\mathrm{1s^2}$ core. 
The numbers of CSFs in the final even and odd state expansions are, respectively, {6 415 798} and {4 988 973}, distributed over the different $J$ symmetries.

\subsection{\ion{C}{III}}
In the computations of beryllium-like carbon, the MR simply consisted of the targeted configurations (see also Table~\ref{tab:MR}). The CSF expansions used in the MCDHF calculations were obtained by allowing SD substitutions from the valence orbitals, accounting for VV correlation effects.
The final wave functions of the targeted states were determined in subsequent
RCI calculations, which included CSFs that were formed by allowing single,
double, and triple (SDT) substitutions from all orbitals of the MR
configurations, with the limitation of leaving no more than one hole in the
$\mathrm{1s^2}$ atomic core. The final even and odd state expansions,
respectively, contained 1 578 620 and 1 274 147 CSFs, distributed over the different $J$ symmetries.

\subsection{\ion{C}{IV}}
Likewise the computations in \ion{C}{III}, the MR in the computations of lithium-like carbon was solely represented by the targeted configurations (see also Table~\ref{tab:MR}). In the MCDHF calculations, the CSF expansions were acquired by implementing SD electron substitutions from the configurations in the MR,
with the restriction of allowing maximum one hole in the $\mathrm{1s^2}$ core. 
In this case, the shape of the correlation orbitals was established by CSFs accounting for valence (V) and CV correlation effects. In the subsequent RCI calculations, the CSF expansions were enlarged by enabling all SDT substitutions from the orbitals in the MR to the active set of orbitals. The final expansions of the atomic states gave rise to 1 077 872 CSFs with even parity and 1 287 706 CSFs with odd parity, respectively, shared among the different $J$ symmetry blocks.

\section{Results}\label{sec:results}

\begin{figure*}
    \centering
    \includegraphics[width=0.48\textwidth,clip]{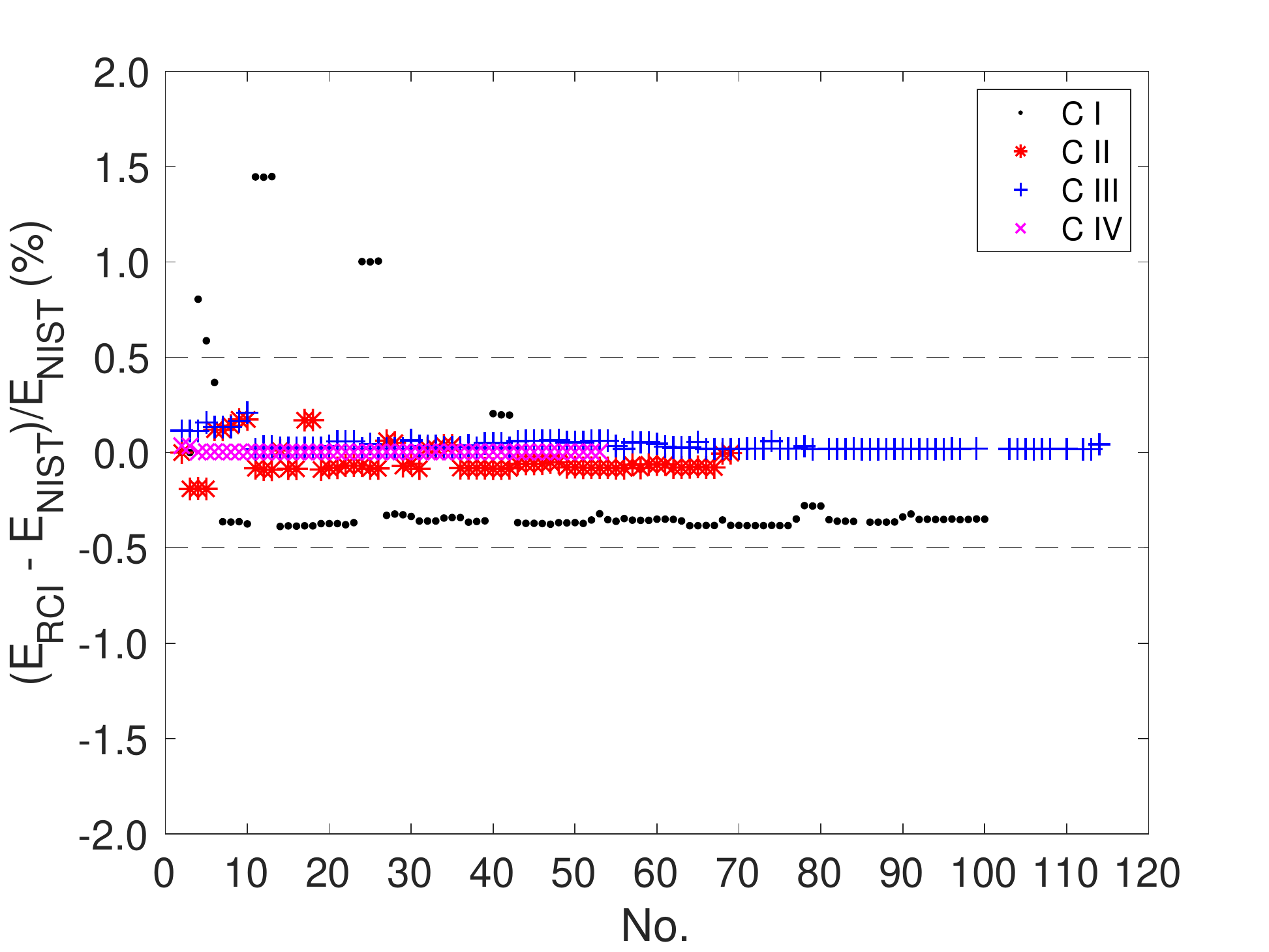}
    \includegraphics[width=0.48\textwidth,clip]{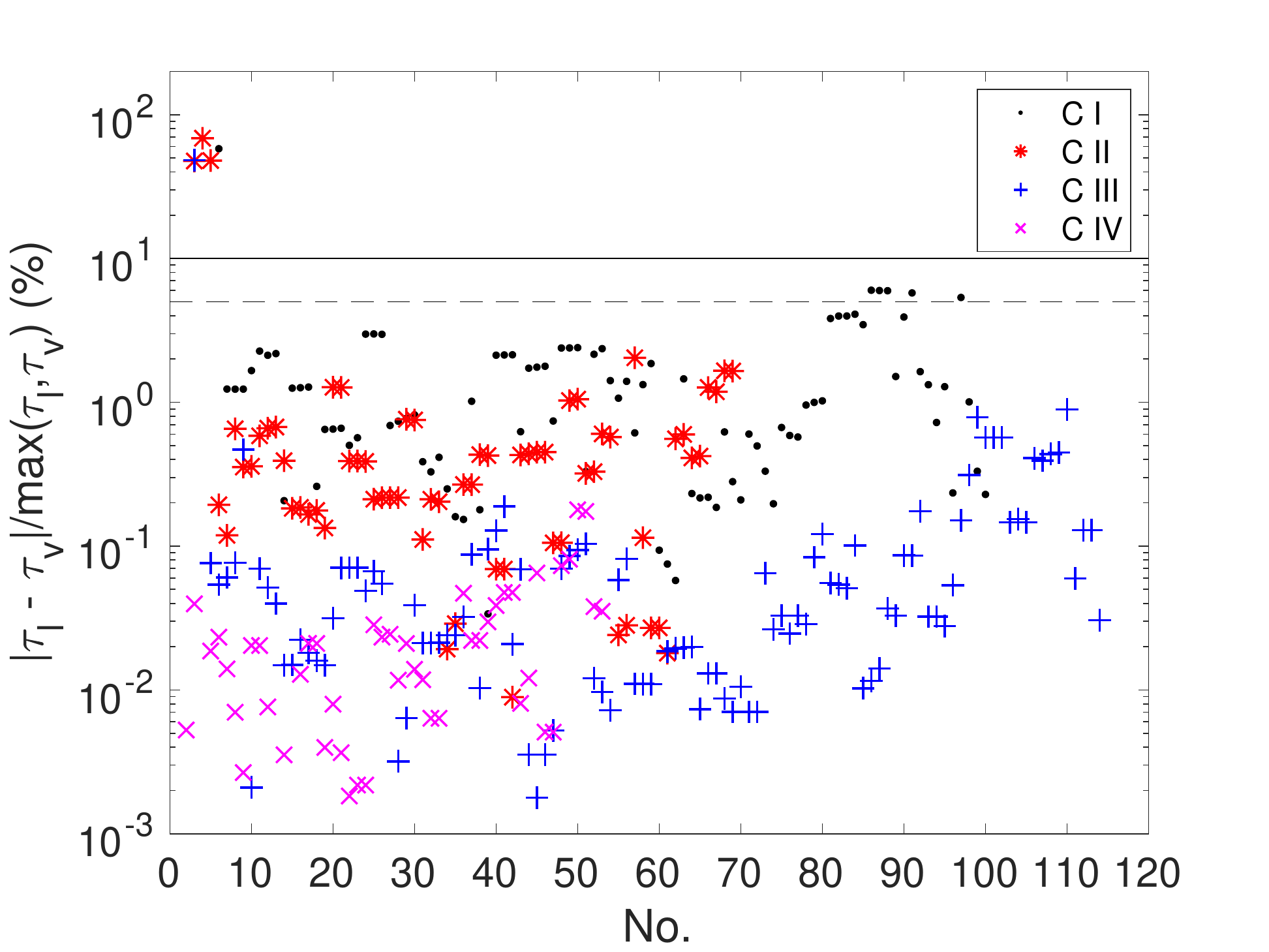}
    \caption{Left panel: Comparison of computed energy levels in the present work with data
    from the NIST database, for \citoiv{}. The dashed lines indicate the $-$0.5\% and
    0.5\% relative discrepancies. Right panel: The relative differences between the lifetimes in length and velocity forms, for \citoiv{}. The dashed and solid lines indicate the 5\% and 10\% relative differences, respectively. No., as label in the x-axis, corresponds to the No. in Table \ref{tab:CI-IV_energy}.}
    \label{fig:lev-tau}
\end{figure*}

The energy spectra and wave function composition in
$LS$-coupling for the  100, 69, 114, and 53 lowest states, respectively,
for \citoiv{} are given in Table \ref{tab:CI-IV_energy}.
In the tables, the states are given with unique labels
\citep{Gaigalas2017}, and the labelling is determined by the CSFs with the largest coefficient in the expansion
of Eq. \eqref{eq:csfs}. 
We first summarise the results here,
before discussing the individual ions 
in detail in Sects. \ref{sec:CI} -- \ref{sec:CIV}, below.

The accuracy of the wave functions 
from the present calculations was 
evaluated by comparing the calculated energy levels with 
experimental data provided via the National Institute of Standards and
Technology (NIST) Atomic Spectra Database \citep{NIST_ASD}. 
In the left panel of Fig. \ref{fig:lev-tau}, energy levels computed in this work
are compared with the NIST data. 
A closer inspection of the figure reveals that the relative discrepancies between the experimental and the computed in this work energies are, in most cases, about $-$0.35\%, $-$0.08\%, 0.03\%, and 0.003\%, respectively, for \citoiv{}. Only for levels of the $\mathrm{2s2p^3}$ configuration in \ion{C}{I}, the
disagreements are larger than 1.0\%.
The average difference of the computed energy levels relative to the energies
from the NIST database is 0.41\%, 0.081\%, 0.041\%, and 0.0044\%, respectively,
for \citoiv{}. 
In Table \ref{tab:CI-IV_energy}, lifetimes in length and
velocity gauges are also presented. The right panel of Fig. \ref{fig:lev-tau} presents the relative
differences between the lifetimes in length and velocity forms for \citoiv{}.
Except for a few long-lived states that can decay to the ground state
only through IC transitions, the relative differences are well below 5\%. 

\setcounter{table}{1}
\begin{table*}
    \centering
    \caption{Distribution of the uncertainties $dT$ (in \%) of the computed transition rates in \citoiv{} depending on the magnitude of the rates. The transition rates are arranged in five groups based on the magnitude of the $A$ values (in s$^{-1}$). The number of transitions, No., the mean $dT$, $\langle dT \rangle$, (in \%), and the standard deviations, $\sigma$, are given for each group of transitions, in \citoiv{}, respectively. The last three rows show the proportions of the transitions with $dT$ less than 20\%, 10\%, and 5\% in all the transitions with $A \ge$ $10^2$ s$^{-1}$ for \ion{C}{I} and \ion{C}{II} and $A \ge$ $10^0$ s$^{-1}$ for \ion{C}{III} and \ion{C}{IV}, respectively.}
    \begin{tabular}{cccccccccccccccccccccc}
    \hline \midrule
    &  \multicolumn{3}{c}{\ion{C}{I}} && \multicolumn{3}{c}{\ion{C}{II}} && \multicolumn{3}{c}{\ion{C}{III}} && \multicolumn{3}{c}{\ion{C}{IV}} \\
     \cmidrule{2-4} \cmidrule{6-8} \cmidrule{10-12} \cmidrule{14-16}
     Group    &  No. & $\langle dT \rangle(\%)$ & $\sigma$ && No. & $\langle dT \rangle(\%)$ & $\sigma$ && No. & $\langle dT \rangle(\%)$ &$\sigma$ && No. & $\langle dT \rangle(\%)$ & $\sigma$\\
     \midrule
 $ < 10^0$  &  62    &       52.6 & 0.34&&    80     &      29.6 &0.32 &&      137   &    10.8 &0.18 &   &  20    &  5.92 &0.061   \\   
 $10^0 - 10^2$ &  156   &       34.0 &0.25 &&    134    &      17.1 &0.24 &&      239   &    5.57 &0.096 &&     10    &  2.38 &0.017   \\
 $10^2 - 10^4$ &  451   &       13.2 &0.15 &&    128    &      14.4 &0.19 &&      354   &    2.48 &0.050 &&     6     &  0.667 &0.0047  \\
 $10^4 - 10^6$ &  600   &       7.20 &0.11 &&    167    &      11.8 &0.15 &&      360   &    1.44 &0.034 &&     43    &  0.267 &0.0035 \\
 $>10^6$   &  284   &       1.68 &0.020 &&    297    &      1.53 &0.023 &&      715   &    0.297 &0.010 &&    307   &  0.205 &0.0041 \\
\midrule
   $dT$ < 20\%      & \multicolumn{3}{c}{87.4\%} && \multicolumn{3}{c}{89.5\%} && \multicolumn{3}{c}{98.4\%} && \multicolumn{3}{c}{100\%} \\
 $dT$ < 10\%      & \multicolumn{3}{c}{77.3\%} && \multicolumn{3}{c}{80.7\%} && \multicolumn{3}{c}{95.7\%} && \multicolumn{3}{c}{100\%} \\
   $dT$ < 5\%       & \multicolumn{3}{c}{62.0\%} && \multicolumn{3}{c}{68.7\%} && \multicolumn{3}{c}{91.7\%} && \multicolumn{3}{c}{99.4\%} \\
\bottomrule
    \end{tabular}
    \label{tab:dT}
\end{table*}

The accuracy of calculated transition rates can be estimated either by 
comparisons with other theoretical works and experimental results, when 
available, or by the quantity $dT$, which is defined in Eq.~\eqref{accuracy} as the
agreement between the values in length and velocity gauges 
\citep{Fischer2009,Ekman2014}. The latter is particularly useful
when no experimental measurements are available.
Transition data, e.g., wavenumbers; wavelengths; line strengths; weighted oscillator
strengths; transition probabilities of E1 transitions; and the
accuracy indicators $dT$, are given in Tables \ref{tab:CI_tr} -- \ref{tab:CIV_tr}, respectively, for \citoiv{}.
Note that the wavenumbers and wavelengths are adjusted to match the level
energy values in the NIST database, which are critically evaluated by \cite{Haris_2017} for \ion{C}{I} and \cite{Moore1993} for \ion{C}{II-IV}. 
When no NIST values are available, the 
wavenumbers and wavelengths are from the present MCDHF/RCI calculations and marked with * in the tables.

To better display the uncertainties $dT$ of the computed transitions rates and their distribution in relation to the
magnitude of the transition rate values $A$, the transitions are organized in five groups based on the magnitude of the $A$ values. A statistical
analysis of the uncertainties $dT$ of the transitions is performed for the
1553, 806, 1805, and 386 E1 transitions, respectively, for \citoiv{}. In Table \ref{tab:dT}, the mean value of the 
uncertainties $\langle dT
\rangle$ and standard deviations $\sigma$ are given for each group of transitions. As seen in Table \ref{tab:dT}, most of the estimated uncertainties $dT$ are well below 10\%.
Most of the strong transitions with $A$ > $10^6$ s$^{-1}$ are associated with small
uncertainties $dT$, less than 2\%, especially for \ion{C}{III} and \ion{C}{IV}, for which $\langle dT
\rangle$ is 0.297\% ($\sigma$ = 0.01) and 0.205\% ($\sigma$ = 0.0041), respectively.
It is worth noting that, by employing the alternative optimization scheme of the radial orbitals in the present calculations, the uncertainties $dT$ for transitions involving high Rydberg states are significantly reduced.

Contrary to the strong
transitions, the weaker transitions are associated with relatively large $dT$ values.
This is even more pronounced for the first two groups of transitions in \ion{C}{I} and \ion{C}{II},
where $A$ is less than $10^2$ s$^{-1}$. 
These weak E1 transitions are either IC or two-electron one-photon
transitions. The rates of the former transitions, in relativistic calculations, are small due to the strong cancellation
contributions to the transition moment \citep{Ynnerman1995}, whereas the rates of 
the latter transitions are identically zero in the simplest approximation of the wave
function and only induced by correlation effects \citep{Bogdanovich_2007,Li_2010}.
These types of transitions are extremely challenging, and therefore interesting from a theoretical point
of view, and improved methodology is needed to further decrease the uncertainties of the respective transition data. 

\begin{figure*}
    \centering
    \includegraphics[width=0.48\textwidth,clip]{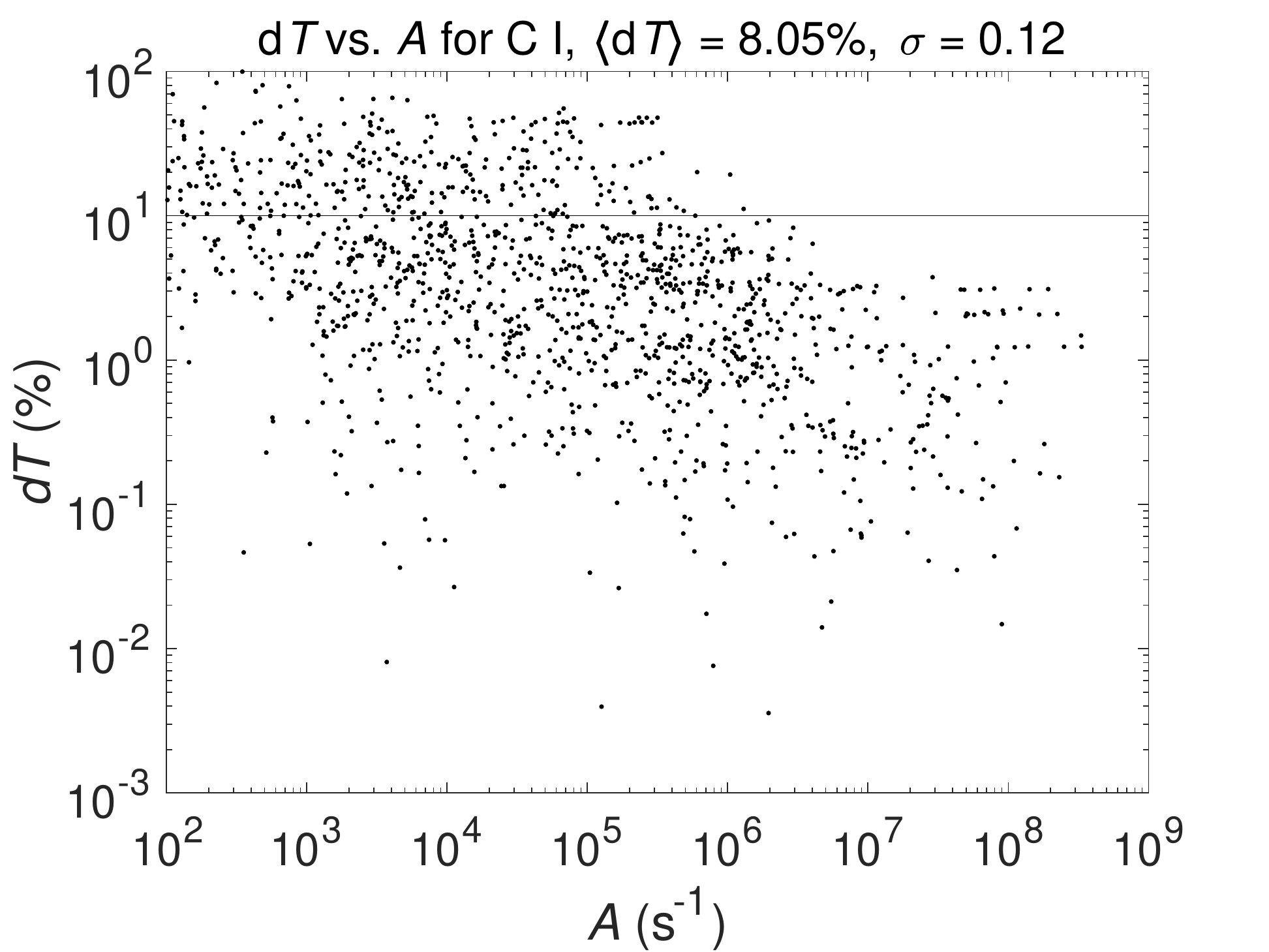}
    \includegraphics[width=0.48\textwidth,clip]{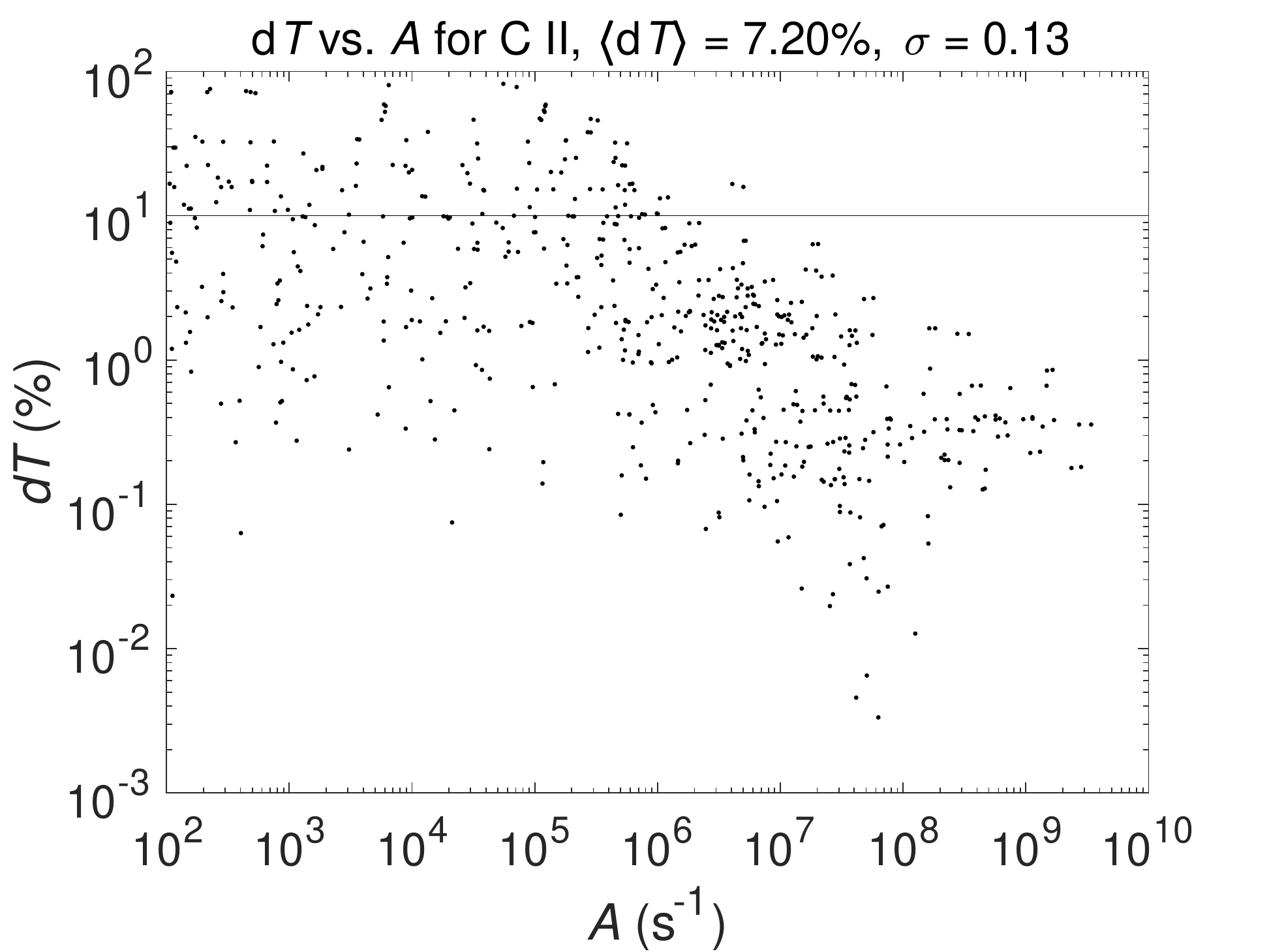}
    \includegraphics[width=0.48\textwidth,clip]{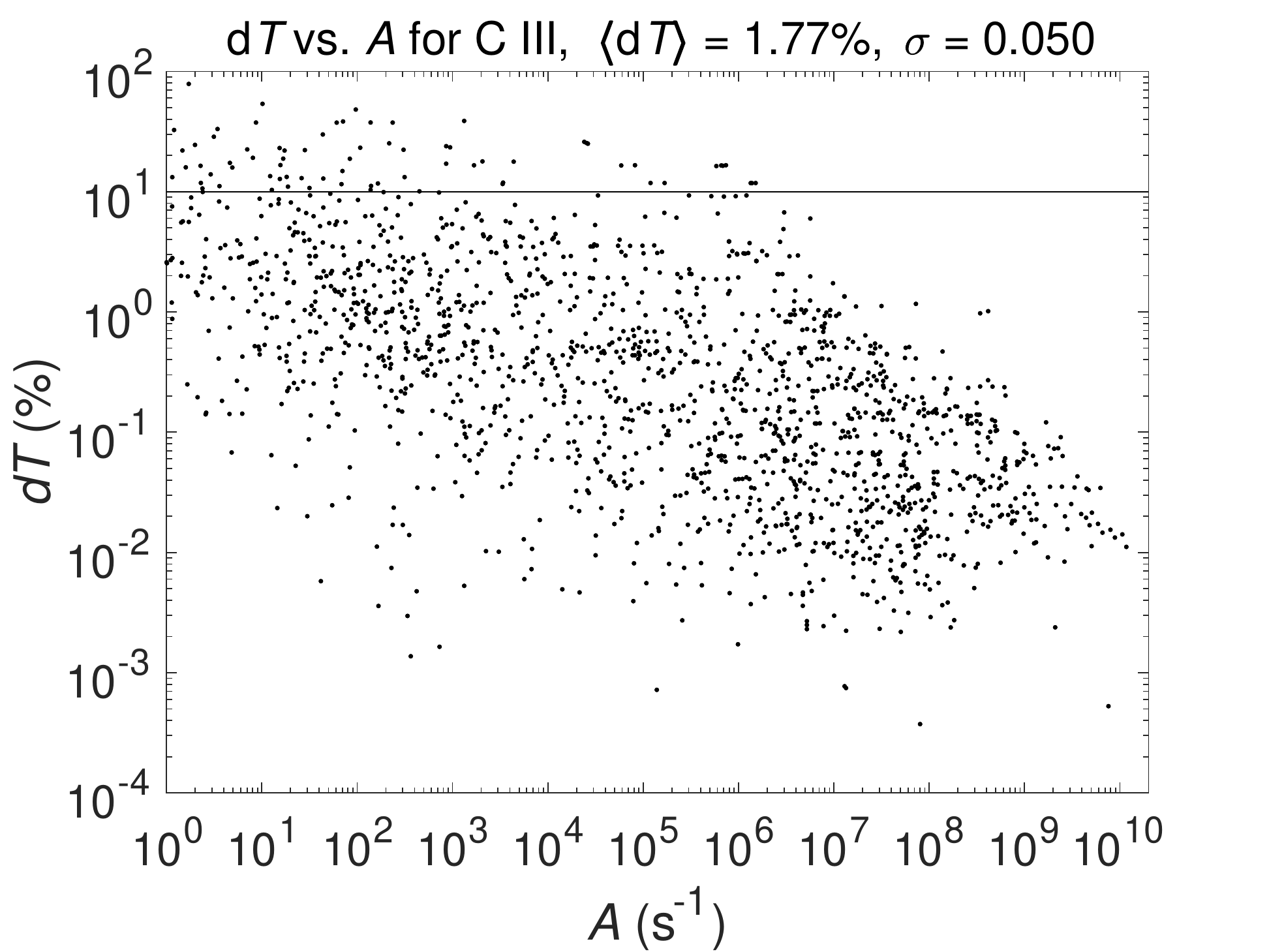}
    \includegraphics[width=0.48\textwidth,clip]{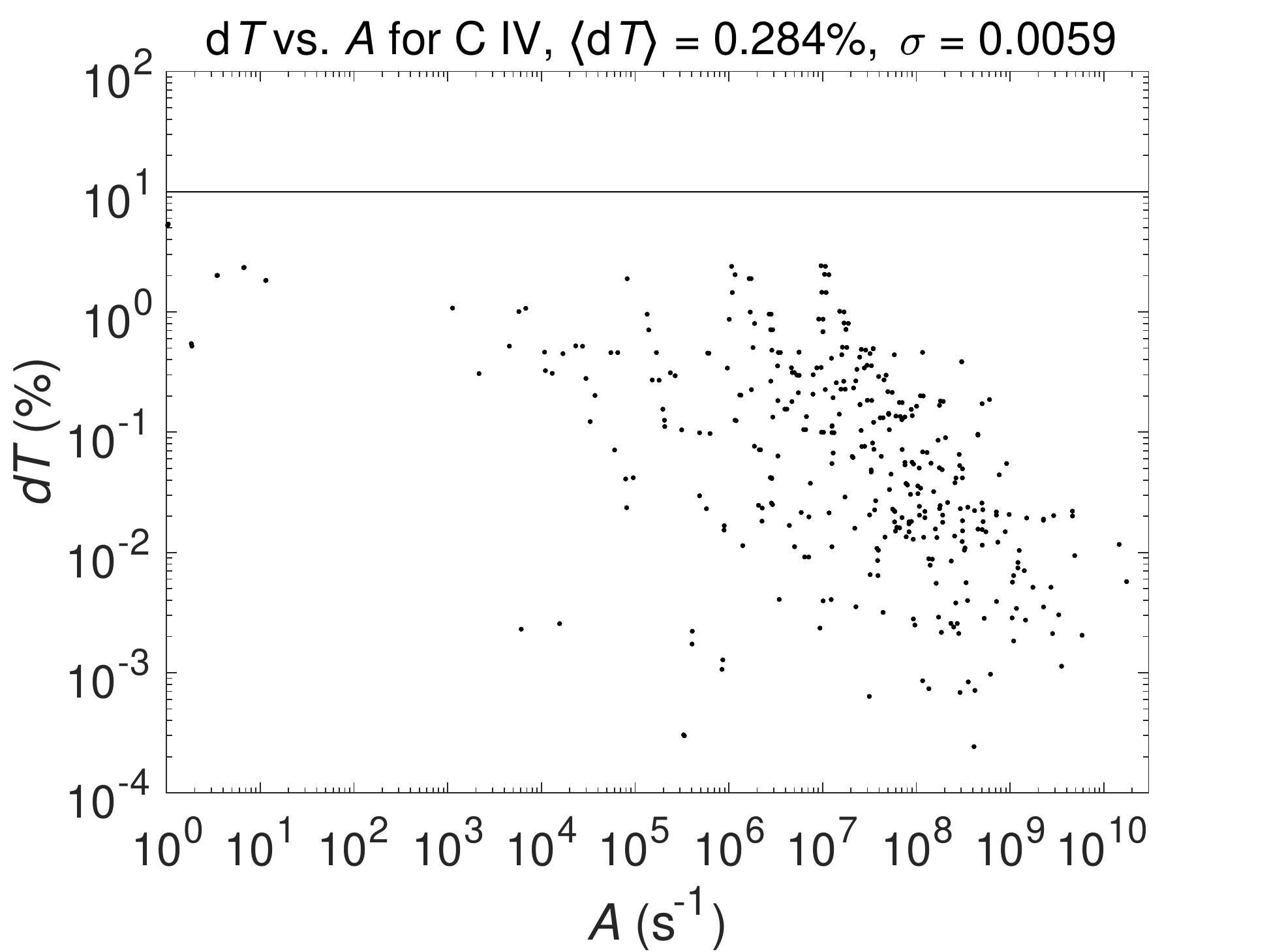}
    \caption{Scatterplot of d$T$ values vs. transition rates $A$ of E1 transitions, for \citoiv{}. 
    The solid lines indicate the 10\% relative agreement between the length and velocity gauges.}
    \label{fig:dT}
\end{figure*}

Fortunately, the weak transitions tend to be of lesser astrophysical
importance, either for opacity calculations, or for spectroscopic
abundance analyses.
Thus, only the transitions with $A \ge$ $10^2$ s$^{-1}$ for \ion{C}{I}
and \ion{C}{II}, and $A \ge$ $10^0$ s$^{-1}$ for \ion{C}{III} and \ion{C}{IV},
are discussed in the paper; although
the complete transition data tables, for all computed E1
transitions in \citoiv{}, are available online.
The scatterplots of $dT$ versus $A$ are given in Fig. \ref{fig:dT}. The mean $dT$ for all presented E1 transitions shown in Fig. \ref{fig:dT} is 8.05\% ($\sigma$ = 0.12), 7.20\% ($\sigma$ = 0.13), 1.77\% ($\sigma$ = 0.05), and 0.28\% ($\sigma$ = 0.0059), respectively, for \citoiv{}. 
A statistical analysis of the proportions of the transitions with $dT$ less than 20\%, 10\%, and 5\% in all the presented E1 transitions is also performed
and shown in the last three rows of Table \ref{tab:dT}.

\begin{figure*}
    \centering
    \includegraphics[width=0.48\textwidth,clip]{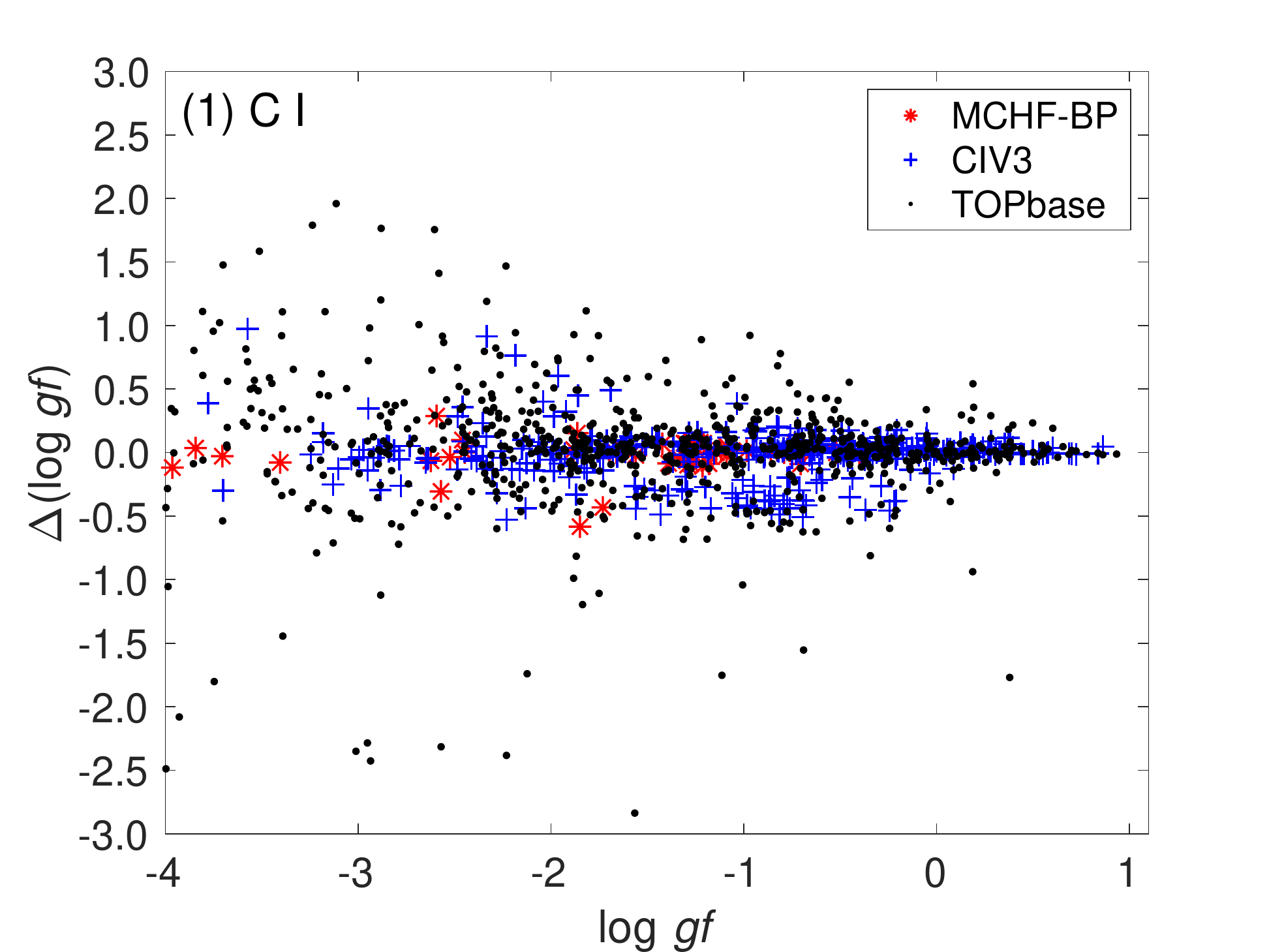}
    \includegraphics[width=0.48\textwidth,clip]{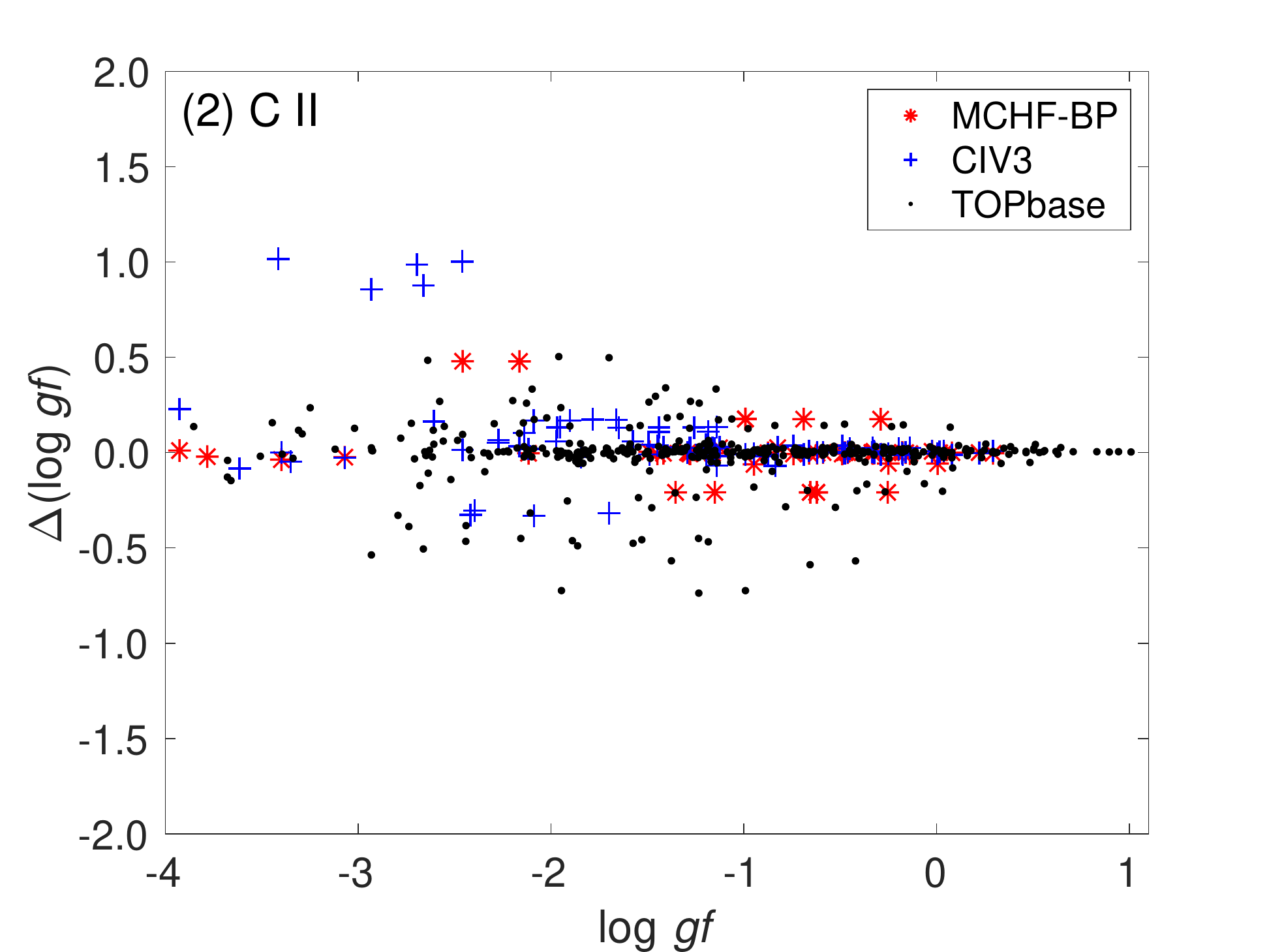}
    \includegraphics[width=0.48\textwidth,clip]{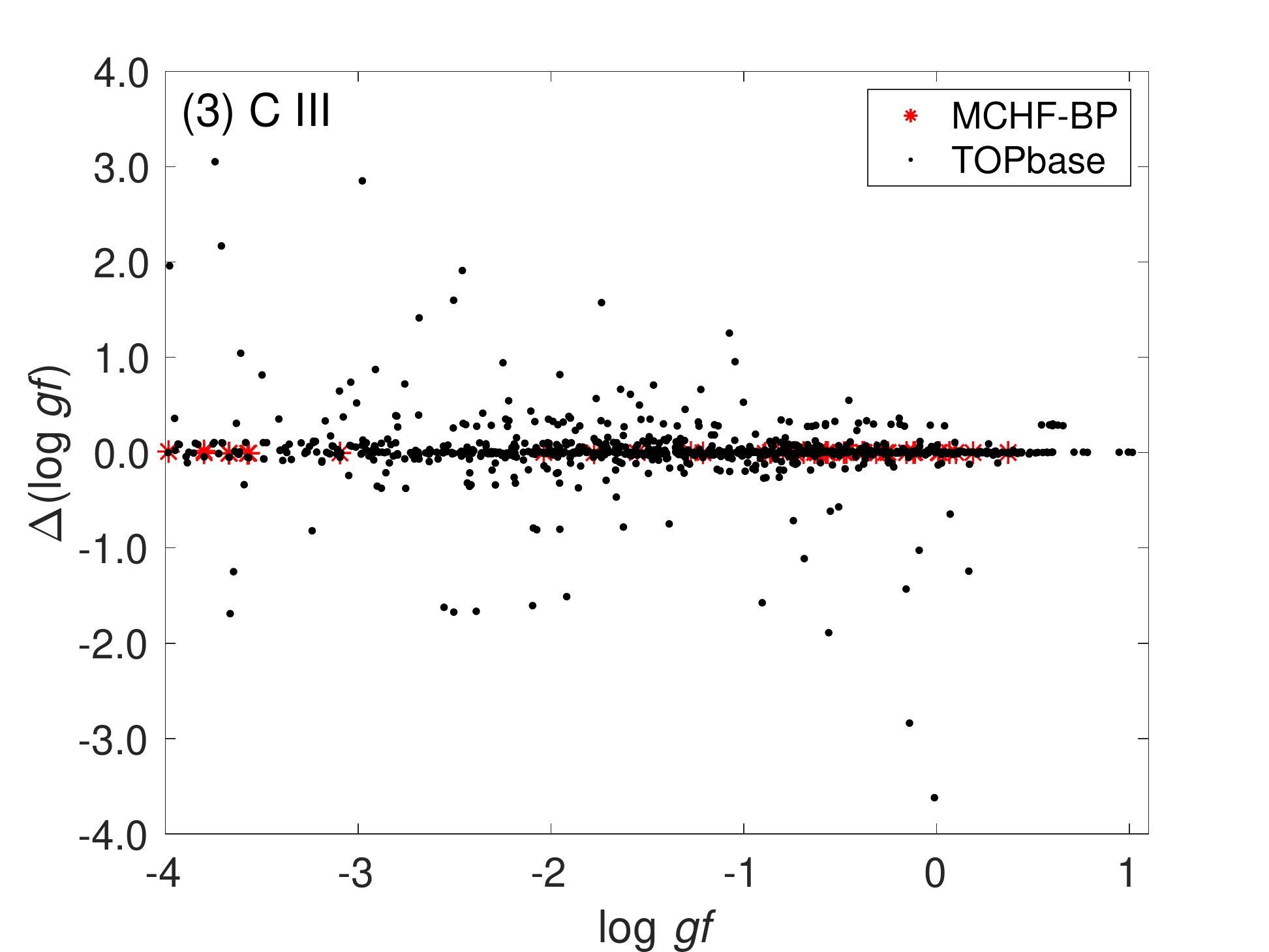}
    \includegraphics[width=0.48\textwidth,clip]{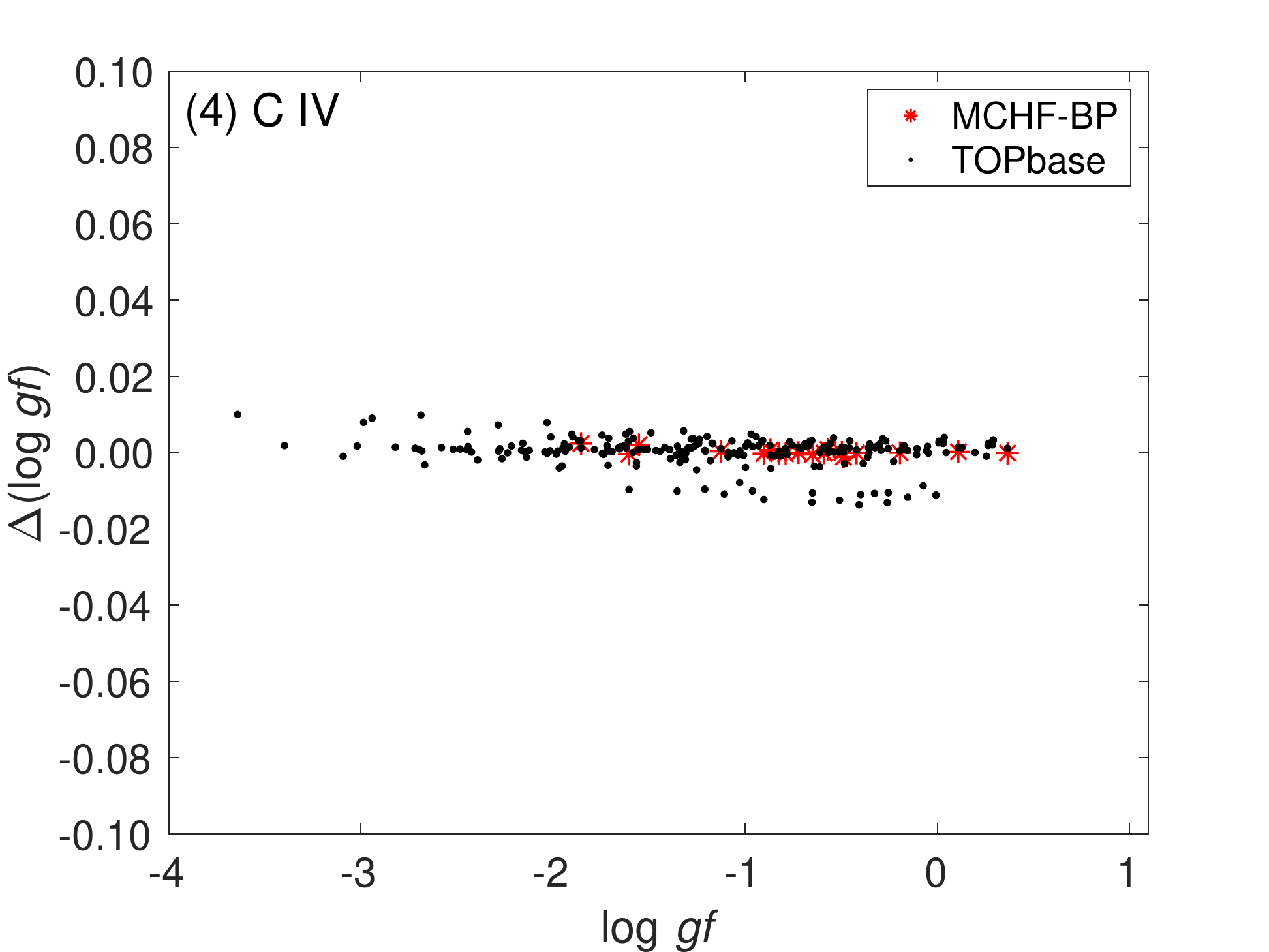}
    \caption{Differences between the calculated $\log gf$ values in this work and results from other theoretical
    calculations: MCHF-BP (red asterisk), CIV3 (blue plus sign), and TOPbase (black point), for \citoiv{}. }
    \label{fig:loggf}
\end{figure*}

Finally, the present work 
can be compared with other theoretical calculations.
In Fig. \ref{fig:loggf}, $\log gf$ values from the present work are compared
with results from MCHF-BP \citep{Fischer_2006,Tachiev_2000,Tachiev_1999,FISCHER1998119}, CIV3 \citep{Hibbert1993,CORREGE200419}, and TOPbase data \citep{1992RMxAA..23..107C}, when available.  As
shown in the figure, the differences between the $\log gf$ values computed in
the present work and respective results from other sources are rather small for most of the transitions.  
Comparing the MCDHF/RCI results with those
from CIV3 calculations by \cite{Hibbert1993}, which are frequently used in the abundance analyses, 292(228) out of 378 transitions are in agreement within 20\% (10\%) for \ion{C}{I}, and 78(66) out of 87 transitions are within the same range for \ion{C}{II}.
The results from the MCDHF/RCI and MCHF-BP calculations are found to be in very good agreement for \ion{C}{III--IV}, with the relative differences being less than 5\% for all the computed transitions. 
More
details about the comparisons with other theoretical calculations, as well as
with experimental results, are given in Sects \ref{sec:CI} -- \ref{sec:CIV}.

\subsection{\ion{C}{I}}\label{sec:CI}
The computed excitation energies, given in Table \ref{tab:CI-IV_energy}, are compared with results from NIST \citep{NIST_ASD}.
With the exception of the
levels belonging to the $\mathrm{2s2p^3}$ configuration, for which the average relative difference between theory and experiment is 1.22\%, the mean relative difference for the rest of the states is 0.35\%.
The complete transition data, for all computed E1 transitions
in \ion{C}{I}, can be found in Table \ref{tab:CI_tr}. Based
on the statistical analysis of the uncertainties $dT$ shown in table \ref{tab:dT}, out of the 1335 transitions with $A \ge$ $10^2$ s$^{-1}$, the proportions of the transitions with $dT$ less than 20\%, 10\%, and 5\% are, respectively, 87.4\%, 77.3\%, and 62.0\%.

In \ion{C}{I}, 
experimental transition data are available for the
$\mathrm{2p3p \rightarrow 2p3s}$, $\mathrm{2p3d \rightarrow 2p3p}$, and
$\mathrm{2p4s \rightarrow 2p3p}$ transition arrays
using a stabilized arc source
\citep{MUSIELOK1997395,Golly_2003,Bacawski_2001}.
In Table~\ref{tab:com_exp}, the experimental relative line strengths, together with their uncertainties, are
compared with the present MCDHF/RCI theoretical values and with values from the non-relativistic CIV3 calculations
by \cite{Hibbert1993} that included semi-empirical diagonal energy shifts by
$LS$ configuration in the interaction matrix in the determination of the
wavefunctions. The estimated uncertainties $dT$ of the MCDHF/RCI line strengths are
given as percentages in parentheses.  
In most cases, the theoretical values fall
into, or only slightly outside, the range of the estimated uncertainties of the
experimental values.  

Comparing the
MCDHF/RCI results with the results from the CIV3 calculations by
\cite{Hibbert1993}, we see that 41 out of the 50 transitions in
common are in good agreement,
with the relative differences being less than 10\%
(see Table \ref{tab:com_exp}).
For the $\mathrm{2p4s~^3P^o \rightarrow 2p3p~^3P}$
transitions and the $\mathrm{2p4s~^3P_2^o \rightarrow 2p3p~^3D_1}$ transition, the $S$ values deduced from the present MCDHF/RCI calculations differ substantially from the experimental values, i.e., by more than 20\%,
while the values from the CIV3 calculations appear to be in better agreement with the
corresponding experimental values.  Based on the agreement between the length and
velocity forms, the estimated uncertainties $dT$ of the present MCDHF/RCI
calculations for the above-mentioned transitions are of the order of 8.5\% and
1.4\%, respectively.  For the $\mathrm{2p3d~^3P_2^o \rightarrow 2p3p~^3P_1}$,
$\mathrm{2p4s~^3P_2^o \rightarrow 2p3p~^3D_2}$, and $\mathrm{2p3d~^3D_2^o
\rightarrow 2p3p~^3D_3}$ transitions, both theoretical results are outside the
range of the estimated uncertainties of the experimental values.  For the
$\mathrm{2p3d~^3D^o \rightarrow 2p3p~^3P}$ transitions, the evaluated relative
line strengths by \cite{Golly_2003} slightly differ from the observations by
\cite{Bacawski_2001}. The latter seem to be in better overall agreement with
the transition rates predicted by the present calculations.

In Table \ref{tab:CI_com}, the computed line strengths and transition rates are
compared with values from the spline frozen-cores (FCS) method by
\cite{Zatsarinny_2002} and the MCHF-BP calculations by \cite{Fischer_2006}.
\cite{Zatsarinny_2002} presented oscillator strengths for transitions from the
$\mathrm{2p^2~^3P}$ term to high-lying excited states, while
\cite{Fischer_2006} considered only transitions from $\mathrm{2p^2~^3P}$,
$\mathrm{^1D}$, and $\mathrm{^1S}$ to odd levels up to $\mathrm{2p3d~^3P^o}$.
As seen in the table, the present MCDHF/RCI results seem to be in better
agreement with the values from spline FCS calculations. 76 out of 98
transitions from \cite{Zatsarinny_2002} agree with present values within 10\%,
while only 38 out of 78 transitions from \cite{Fischer_2006} are within the
same range. The relatively large differences with \cite{Fischer_2006} may be
due to the fact that limited electron correlations were included in
their
calculations. In the MCHF-BP calculations, two types of correlation, i.e., VV,
CV, have been accounted for; however, the CC correlation has not been
considered. Additionally, CSF expansions obtained from SD substitutions are not
as large as the CSF expansions used in the present calculations.  For the majority of the strong
transitions with $A$ > $10^6$ s$^{-1}$, there is a very good agreement between
the MCDHF/RCI results and the spline FCS values, with the relative difference
being less than 5\%. On the other hand, for the $\mathrm{2p3d~^3F \rightarrow
2p^2~^3P}$ and $\mathrm{2p4s~^1P_1 \rightarrow 2p^2~^3P}$ transitions, the
observed discrepancies between these three methods, i.e., MCDHF/RCI, spline FCS, and MCHF-BP, are quite large. These transitions are all $LS$-forbidden transitions, the former is with
$\Delta L$ = 2 and the latter is spin-forbidden transition; these types of transitions are challenging for computations and are 
always with large uncertainties. For example, for the $\mathrm{2p3d~^3F_3 \rightarrow
2p^2~^3P_2}$ transition, the $A$ values from MCDHF/RCI, spline FCS, and MCHF-BP calculations are, respectively, 7.92E+06, 6.24E+06, and 1.14E+07 s$^{-1}$, with the relative difference between each two of them being greater than 20\%. Experimental data are, therefore, needed for validating these theoretical results. On the contrary, based on the agreement between the
length and velocity forms displayed in the parentheses, the estimated uncertainties of the MCDHF/RCI
calculations for the above-mentioned transitions are all less than 0.5\%.

\subsection{\ion{C}{II}}\label{sec:CII}
The relative differences between theory and experiment 
for all the energy levels of $\mathrm{2s2p^2}$ are 0.16\%, while
the mean relative difference for the rest of the states is 0.071\% (see Table~\ref{tab:CI-IV_energy}). 
The complete transition data, for all computed E1 transitions 
in \ion{C}{II}, can be found in Table \ref{tab:CII_tr}. 
Out of the presented 592 E1 transitions with $A \ge$ $10^2$ s$^{-1}$, the proportions of the transitions with $dT$ less than 20\%, 10\%, and 5\% are, respectively, 89.5\%, 80.7\%, and 68.7\%.

In Table \ref{tab:com_exp}, the lifetimes from the present MCDHF/RCI
calculations are compared with available results from the MCHF-BP calculations
by \cite{Tachiev_2000} and observations by \cite{Reistad_1986} and \cite{Trabert_1999}.
\cite{Trabert_1999} measured lifetimes for the three fine-structure components
of the $\mathrm{2s2p^2~^4P}$ term in an ion storage ring. For the measured
lifetimes by \cite{Reistad_1986} of the doublets terms using the beam-foil
technique, a single value for the two fine-structure levels is provided.
It can be seen that, in all cases, the MCDHF/RCI computed lifetimes agree with the experimental values by \cite{Reistad_1986} within the experimental errors. 
For the $\mathrm{2s2p^2~^4P_{1/2,3/2,5/2}}$
states,
as discussed in Sect.~\ref{sec:intro}, the discrepancies between the measured 
transition rates by \cite{Fang1993} and by \cite{Trabert_1999} are quite large.
It is found that the MCDHF/RCI values are in better agreement with the results given by the latter measurements, with a
relative difference less than 3\%.
For these long-lived states, 
the measured lifetimes are better
represented by the MCDHF/RCI results than by the MCHF-BP values.

The computed line strengths and transition rates are compared with values from
the MCHF-BP calculations by \cite{Tachiev_2000} and the CIV3 calculations by
\cite{CORREGE200419} in Table \ref{tab:CII_com}. We note that the agreement
between the present MCDHF/RCI and the MCHF-BP transition rates exhibits a broad
variation. In the earlier MCHF-BP and our MCDHF/RCI calculations, the same correlation
effects, i.e., VV and CV, have been accounted for. However, the CSF expansions
obtained from SD substitutions in the MCHF-BP calculations are not as large as
the CSF expansions used in the present calculations, and as a consequence, the
$LS$-composition of the configurations might not be predicted as accurately in
the former calculations. The MCDHF/RCI results seem to be in better overall
agreement with the values from the CIV3 calculations, except for transitions from
$\mathrm{2p^3~^2P^o}$ to $\mathrm{2s2p^2~\{^4P, {^2}S\}}$ and to
$\mathrm{2s^23d~^2D}$. For these transitions, involving $\mathrm{2p^3~^2P^o}$
as the upper level, the transition rates $A$ are of the order of $10^2$ --
$10^4$ s$^{-1}$. The $dT$ values are relatively large in the present calculations. This is
due to the strong cancellation effects caused by, e.g., the strong 
mixing between the $\mathrm{2p^3~^2P^o}$ and $\mathrm{2s2p3s~^2P^o}$ levels 
for $\mathrm{2p^3~^2P^o} \rightarrow \mathrm{2s2p^2~{^2}S}$, and the mixing between 
the $\mathrm{2p^3~^2P^o}$ and $\mathrm{2s^24p~^2P^o}$ levels for 
$\mathrm{2p^3~^2P^o} \rightarrow \mathrm{2s^23d~{^2}D}$. 
Large discrepancies are also observed between the MCDHF/RCI and MCHF-BP results, as
well as between the MCHF-BP and CIV3 results for these transitions. 
Experimental data are, therefore, crucial for validating the aforementioned theoretical results.
On the contrary, for the majority of the strong
transitions with $A$ > $10^6$ s$^{-1}$, there is a very good agreement between
the MCDHF/RCI results and those from the two previous calculations, with the
relative differences being less than 5\%.

\subsection{\ion{C}{III}}\label{sec:CIII}
The average relative discrepancy between the computed excitation energies, shown in Table~\ref{tab:CI-IV_energy}, and the NIST recommended values is 0.041\%.
The complete transition data, for all computed E1 transitions
in \ion{C}{III}, can be found in Table \ref{tab:CIII_tr}. Out of the 1668 transitions with $A \ge$ $10^0$ s$^{-1}$, 91.7\% (98.4\%) of them 
have $dT$ values less than $5\%$ (20\%). Further, the mean $dT$ for all transitions with $A \ge$ $10^0$ s$^{-1}$ is 1.8\% with $\sigma$ = 0.05.

The lifetimes of the $\mathrm{2s2p~^1P^o_1}$, $\mathrm{2p^2~\{^1S_0, ^1D_2\}}$, and $\mathrm{2s3s~^1S_0}$ states were measured by \cite{Reistad_1986} using the beam-foil technique,
and the oscillator strengths for the $\mathrm{2s2p~^1P^o_1 \rightarrow 2s^2~^1S_0}$ and the $\mathrm{2p^2~\{^1S_0, ^1D_2\} \rightarrow 2s2p~^1P^o_1}$ transitions were also provided.
Table \ref{tab:com_exp} gives the comparisons between the observed and computed oscillator strengths and lifetimes
in \ion{C}{III}. Looking at the table, we see an excellent agreement between the present calculations and those from the MCHF-BP calculations \citep{Tachiev_1999} with the relative difference being less than 0.7\%. In all cases, the computed oscillator strengths and lifetimes agree with experiment within the experimental errors. 
The exceptions are the oscillator strength of the $\mathrm{2p^2~^1S_0 \rightarrow 2s2p~^1P^o_1}$ transition and the lifetime of 
the $\mathrm{2p^2~^1S_0}$ state, for which the computed values slightly differ from the observations.

In Table \ref{tab:CIII_com}, the computed line strengths and transition rates are compared with values from the MCHF-BP calculations by \cite{Tachiev_1999} and the \textsc{Grasp} calculations by \cite{Aggarwal_2015}. 
For the majority of the transitions, there is an excellent agreement between the MCDHF/RCI and MCHF-BP values with the relative differences being less than 1\%. Only 4 out of 60 transitions display discrepancies that are greater than 20\%. 
These large discrepancies are observed for the IC transitions, e.g., $\mathrm{2s3d~^3D_2 \rightarrow 2s2p~^1P^o_1}$ and 
$\mathrm{2s3d~^3D_2 \rightarrow 2s2p~^1P^o_1}$, for which the $dT$ is relatively large. 
The discrepancies between the MCDHF/RCI and \textsc{Grasp} values are overall large;
this is due to the fact that limited electron correlations were included in their
calculations. Based on the excellent agreement between the 
MCDHF/RCI and MCHF-BP results as well as with experiment, we believe that
the present transition rates together with the MCHF-BP transition data are more reliable than the ones provided by \cite{Aggarwal_2015}.

\subsection{\ion{C}{IV}}\label{sec:CIV}
The mean relative discrepancy between the computed excitation energies, given in Table~\ref{tab:CI-IV_energy}, and the NIST values is 0.0044\%.
Out of the presented 366 transitions with $A \ge$ $10^0$ s$^{-1}$ shown in Table \ref{tab:CIV_tr}, only two of them have $dT$ values greater than $5\%$; 94.0\% of them with $dT$ being less than 1\%. The mean $dT$ for all transitions with $A \ge$ $10^0$ s$^{-1}$ is $0.28\%$ with $\sigma$ = 0.0059.

For \ion{C}{IV}, there are a number of measurements of transition properties. The transition rates of the $\mathrm{2p~^2P^o_{1/2,3/2} \rightarrow 2s~^2S_{1/2}}$ transitions 
were measured by \cite{KNYSTAUTAS197175} using the beam-foil technique. By
using the same technique, the lifetimes for a number of excited states were measured in four different experimental works 
\citep{Donnelly1978,Buchet1973,Jacques1980,Peach1988}.
In Table \ref{tab:com_exp}, we compare the theoretical results, from present calculations and MCHF-BP calculations, with the NIST recommended values and observed values. The transition rates of the $\mathrm{2p~^2P^o_{1/2,3/2} \rightarrow 2s~^2S_{1/2}}$ transitions from the present work agree perfectly with the values from the MCHF-BP calculations by \cite{FISCHER1998119}, while they are slightly smaller than the NIST data and the values by \cite{KNYSTAUTAS197175}. 
A comparison of the lifetimes of the $\mathrm{\{3s, 4s, 2p, 3p, 4p, 3d, 4d, 5d\}}$ states is made with other theoretical results, i.e., from the MCHF-BP calculations
and the Model Potential method. The agreements between these different theoretical results are better than 1\% for all these states.
Furthermore, the agreement between the computed values and those from observations is also very good except for the $\mathrm{3s~^2S_{1/2}}$ level,
for which the MCDHF/RCI calculations give a slightly smaller lifetime of 0.2350~ns than the observed value of 0.25 $\pm$ 0.01 ns.

In Table \ref{tab:CIV_com}, the computed line strengths and transition rates are compared with available values from the MCHF-BP calculations by \cite{FISCHER1998119}. There is an excellent agreement between the two methods with the relative differences being less than 1\% for all transitions.

\section{Reanalysis of the solar carbon abundance}
\label{solaranalysis}

\setcounter{table}{2}
\begin{table*}
\begin{center}
    \caption{The $14$ permitted \ion{C}{I} lines used 
as abundance diagnostics in \citet{2019A&A...624A.111A}.
Shown are the upper and lower configurations, 
oscillator strengths 
obtained from the present calculations,
and oscillator strengths from NIST; the latter being based on 
the calculations from CIV3 \citep{Hibbert1993}.
The estimated uncertainties $dT$ of the oscillator strengths are given as
percentages in parentheses.
The final two columns show the abundances 
derived in \citet{2019A&A...624A.111A}, and the post-corrected
values derived here based on the formula $\Delta\log\mathrm{\upvarepsilon_C}^{\text{line}}=
    -\Delta\log gf^{\text{line}}$.}
\label{tab:dcabund}
\begin{tabular}{l l c c c c c c c c}
\hline\midrule
& & & \multicolumn{3}{c}{$\log gf$} \\
\cmidrule{4-6}
     Upper        & Lower             &    $\lambda_\text{air}$(nm) & 
     NIST && MCDHF/RCI($dT$)
     && $\log\upvarepsilon_{\mathrm{C}}^{\text{A19}}$ & 
     $\log\upvarepsilon_{\mathrm{C}}^{\text{L20}}$ \\ 
\midrule
$\mathrm{2p4p~^1D_{2}}$    & $\mathrm{2p3s~^1P_{1}^o}$  &   505.217   & 
    -1.30 &&  
    -1.36(0.8\%)    & &
    8.41 & 8.47\\
$\mathrm{2p4p~^1P_{1}}$    & $\mathrm{2p3s~^1P_{1}^o}$  &   538.034   &
    -1.62 && 
    -1.71(1.4\%)    & & 
    8.43 & 8.52\\
$\mathrm{2p4d~^1P_{1}^o}$  & $\mathrm{2p3p~^1P_{1}}$    &   658.761   &
    -1.00 && 
    -1.05(0.2\%)    & & 
    8.33 & 8.38\\
$\mathrm{2p4d~^3F_{2}^o}$  & $\mathrm{2p3p~^3D_{1}}$    &   711.148   &
    -1.08 && 
    -1.24(0.9\%)    & & 
    8.31 & 8.47\\
$\mathrm{2p4d~^3F_{4}^o}$  & $\mathrm{2p3p~^3D_{3}}$    &   711.318   &
    -0.77 && 
    -0.94(1.5\%)    & & 
    8.41 & 8.58\\
$\mathrm{2p3p~^3D_{1}}$    & $\mathrm{2p3s~^3P_{2}^o}$  &   1075.40   &
    -1.61 && 
    -1.62(1.3\%)    & & 
    8.49 & 8.50\\
$\mathrm{2p3d~^3F_{2}^o}$  & $\mathrm{2p3p~^3D_{2}}$    &   1177.75   &
    -0.52 && 
    -0.46(0.9\%)    & & 
    8.46 & 8.40\\
$\mathrm{2p3d~^3P_{1}^o}$  & $\mathrm{2p3p~^3P_{0}}$    &   1254.95   &
    -0.57 && 
    -0.65(3.3\%)    & & 
    8.51 & 8.60\\
$\mathrm{2p3d~^3P_{0}^o}$  & $\mathrm{2p3p~^3P_{1}}$    &   1256.21   &
    -0.52 && 
    -0.61(3.3\%)    & & 
    8.51 & 8.60\\
$\mathrm{2p3d~^3P_{1}^o}$  & $\mathrm{2p3p~^3P_{1}}$    &   1256.90   &
    -0.60 && 
    -0.70(3.2\%)    & & 
    8.46 & 8.56\\
$\mathrm{2p3d~^3P_{2}^o}$  & $\mathrm{2p3p~^3P_{1}}$    &   1258.16   &
    -0.54 && 
    -0.61(3.4\%)    & & 
    8.46 & 8.53\\
$\mathrm{2p3d~^1P_{1}^o}$  & $\mathrm{2p3p~^1S_{0}}$    &   2102.31   &
    -0.40 && 
    -0.39(0.5\%)    & & 
    8.47 & 8.46\\
$\mathrm{2p4p~^1D_{2}}$    & $\mathrm{2p3d~^1F_{3}^o}$  &   3085.46   &
    +0.10 && 
    +0.07(0.2\%)    & & 
    8.41 & 8.44\\
$\mathrm{2p4d~^1D_{2}^o}$  & $\mathrm{2p4p~^1P_{1}}$    &   3406.58   &
    +0.44 && 
    +0.45(3.1\%)    & & 
    8.47 & 8.46\\
\hline
\end{tabular}
\end{center}
\end{table*}

\begin{figure}
    \includegraphics[scale=0.31]{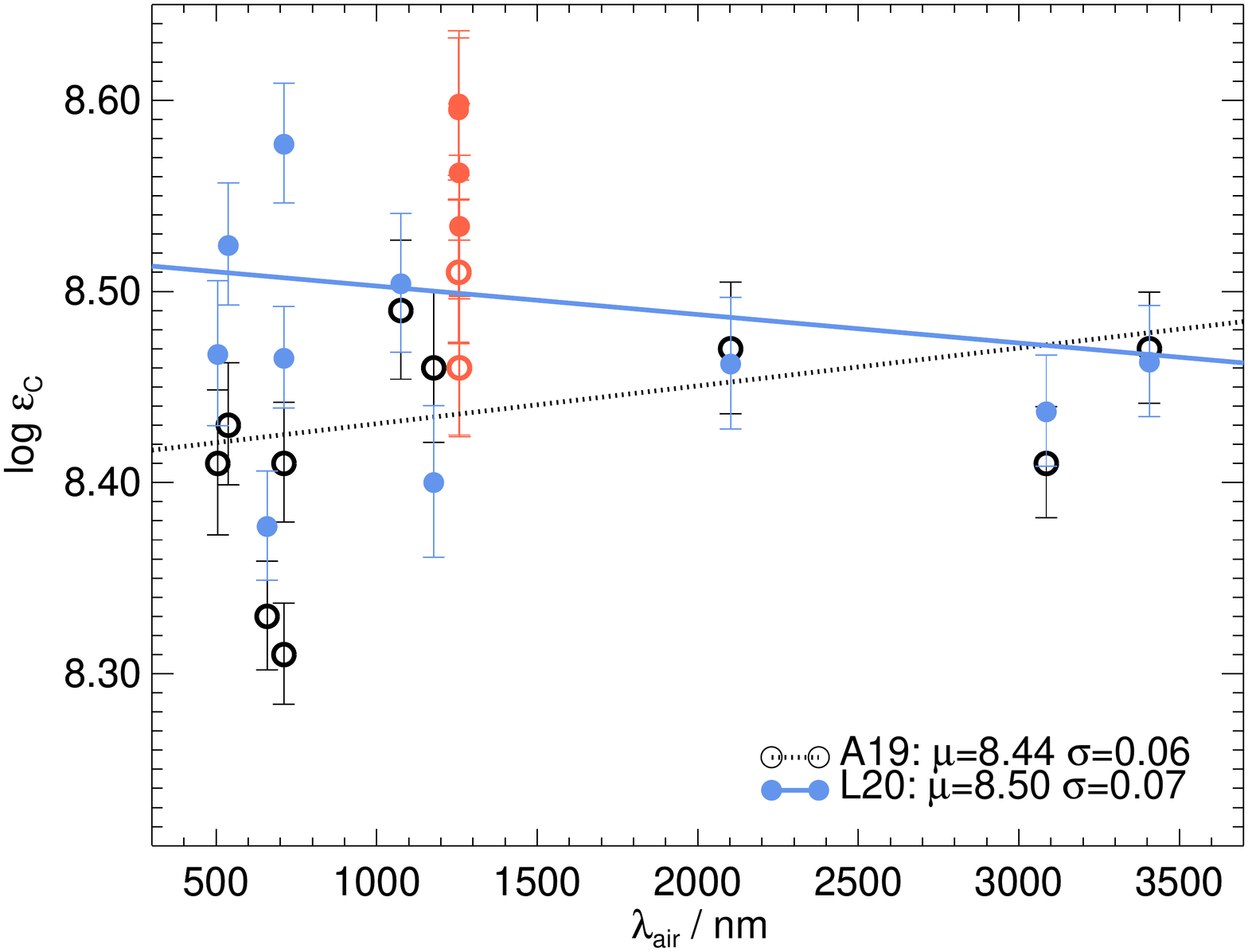}
    \includegraphics[scale=0.31]{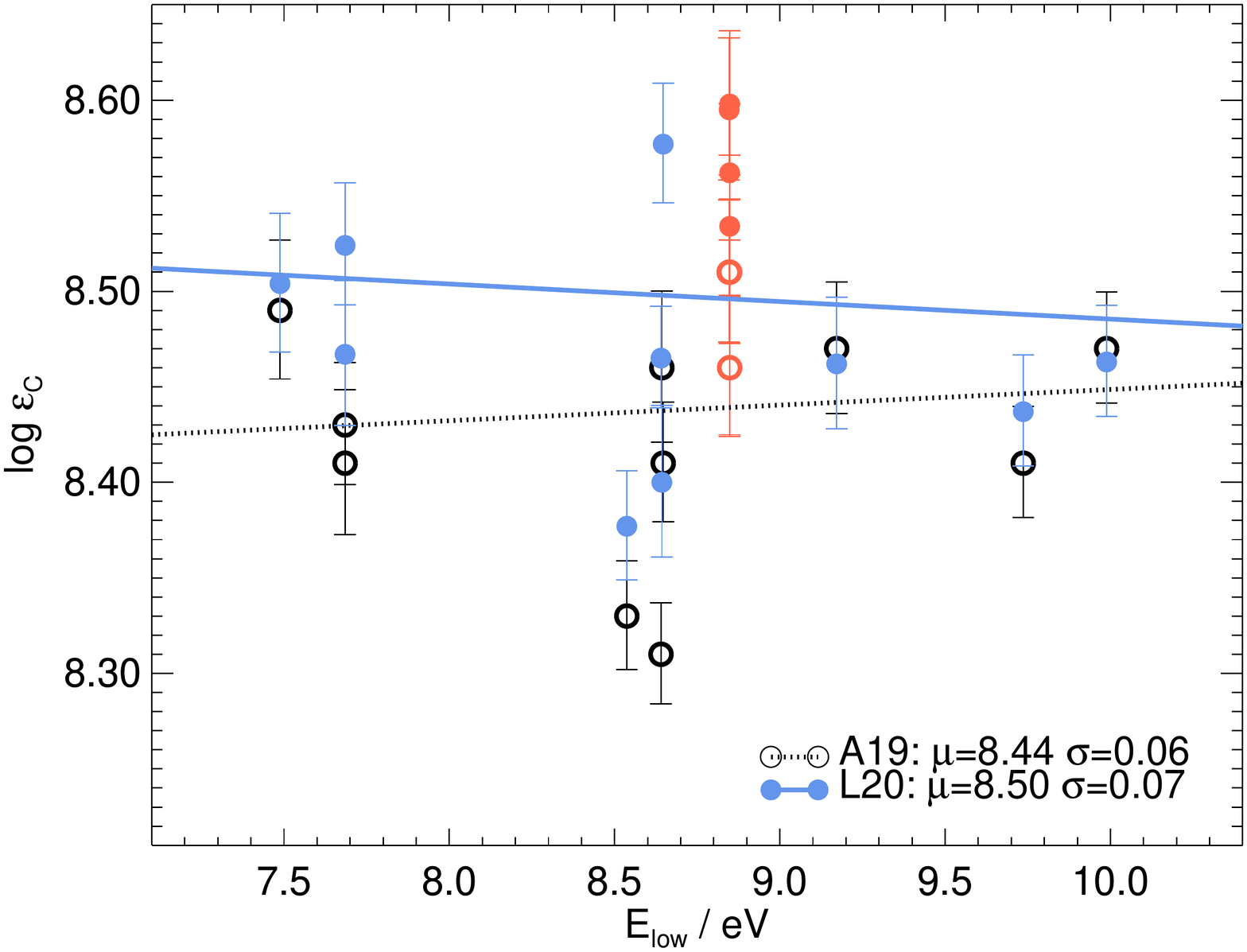}
    \includegraphics[scale=0.31]{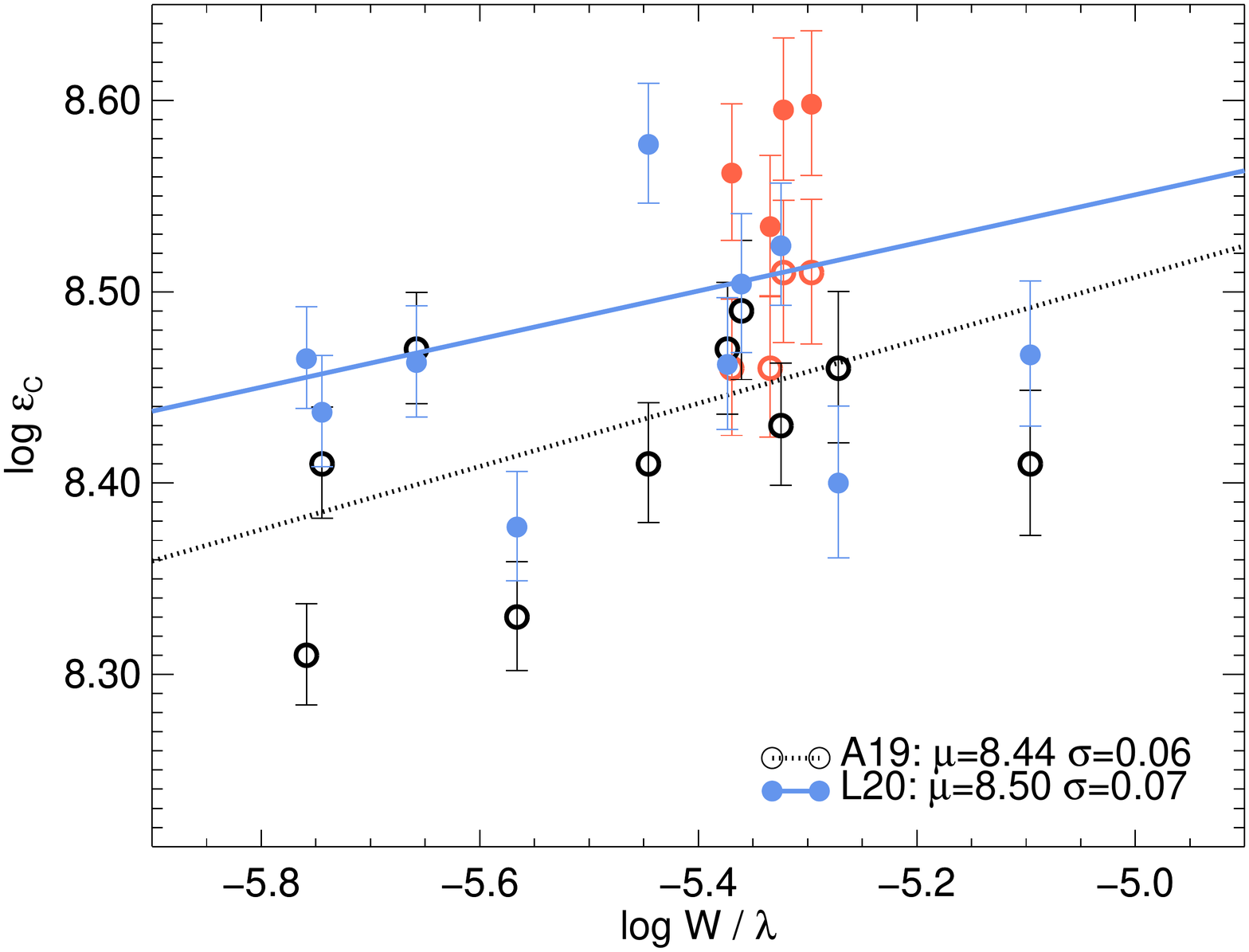}
    \caption{Inferred solar carbon abundances.
    Black points (A19) are the 3D non-LTE results of 
    \citet{2019A&A...624A.111A} for $14$ permitted
    \ion{C}{I} lines.
    Blue points (L20) are these same results but post-corrected
    using the new $\log gf$ data.
    Error bars reflect $\pm5\%$ uncertainties in the measured
    equivalent widths as stipulated by those authors.
    The four lines between
    $1254\,\mathrm{nm}$ and $1259\,\mathrm{nm}$ discussed in the text
    have been highlighted in red.
    The unweighted means $\mu$ (including all $14$ lines)
    and the standard deviations of the samples
    $\sigma$ are stated in each panel.}
    \label{fig:solarcarbon}
\end{figure}

One can also attempt to verify the present atomic data
empirically, in an astrophysical context.
To demonstrate this,
a solar carbon abundance analysis was carried out,
based on permitted \ion{C}{I} lines.
Larger errors in the atomic data usually impart a
larger dispersion in the line-by-line abundance results,
as well as trends in the results with respect to the line 
parameters.

The solar carbon abundance analysis
recently presented in \citet{2019A&A...624A.111A}
was taken as the starting point.
Their analysis is based on
equivalent widths measured in the solar disk-centre
intensity, for $14$ permitted \ion{C}{I} lines
in the optical and near-infrared, 
as well as a single forbidden [\ion{C}{I}] line
at $872.7\,\mathrm{nm}$.
Their analysis draws on 
a three-dimensional (3D) hydrodynamic model solar atmosphere
and 3D non-local thermodynamic equilibrium (non-LTE) radiative transfer,
that reflects the current state-of-the-art
in stellar elemental abundance determinations
\citep[e.g.][]{2009ARA&amp;A..47..481A}.
For the $14$ permitted \ion{C}{I} lines,
the authors adopted transition probabilities from
NIST, that are based on those of
\citet{Hibbert1993} but normalized to 
a different scale \citep{Haris_2017}, corresponding
to differences of the order $\pm0.01\,\dex{}$.

Here, we post-correct the solar carbon abundances 
inferred in \citet{2019A&A...624A.111A} 
from the $14$ permitted \ion{C}{I} lines,
using the new 
atomic data derived in the present study (see Table~\ref{tab:dcabund}).
To first-order, for a given spectral line,
the change in the inferred abundances are related to the 
difference in the adopted transition probabilities
simply as $\Delta\log\upvarepsilon_{\mathrm{C}}^{\text{line}}=
-\Delta\log gf^{\text{line}}$.
We briefly note that second-order effects on the inferred abundances,
propagated forward from changes to the non-LTE statistical equilibrium
when adopting the full set of new $\log gf$ data in the non-LTE 
model atom, were also tested; these were found to be negligible.

The results of this post-correction are illustrated in
Fig. \ref{fig:solarcarbon}.
We find that the dispersion in the line-by-line abundance
results are similar when using the new and the old sets
of $\log gf$ data. 
We also find that the trends in the results with respect to 
the line parameters are of similar gradients.
This is consistent with the finding in
Sect.~\ref{sec:CI}, that the precision of this 
new, much larger atomic data 
set is comparable to that of \citet{Hibbert1993}.

This new analysis implies a solar carbon abundance
of $8.50\,\dex{}$, which is
$0.06\,\dex{}$ larger than that inferred 
in \citet{2019A&A...624A.111A} from \ion{C}{I} lines,
and $0.07\,\dex{}$ larger than the current standard value from 
\citet{2009ARA&amp;A..47..481A} that is based on \ion{C}{I} lines
as well as on molecular diagnostics.
This increase in the mean abundance is due to $12$ of the $14$ 
permitted \ion{C}{I} lines having lower oscillator strengths 
in the present calculations, compared to the NIST data set. 
Six of the lines give results that are larger than the mean ($\log\upvarepsilon\geq8.51$); 
included in this set are all four of the lines between $1254\,\nm$ and $1259\,\nm$, 
which give rise to values of between $8.53$ and $8.60\,\dex$.  
These four lines have the same upper level configuration, 
$\mathrm{2p3d\,^{3}P^{o}}$, and a closer inspection of the $LS$-composition 
reveals that these states are strongly mixed (of the order of $26$\%) with 
$\mathrm{2s2p^3\,^{3}P^{o}}$ states, which are less accurately described in the present calculations. 
As a consequence, as shown in Table~\ref{tab:dcabund}, these transitions appear to be 
associated with slightly larger uncertainties $dT$ than most of the other lines.  
Omitting these four lines, or adopting NIST oscillator strengths for them, 
would reduce the mean abundance from $8.50$ to $8.47\,\dex$.

Given that the scatter and trends in the results
do not support one set of data over the other,
we refrain from advocating a higher solar carbon
abundance at this point.
Nevertheless, this quite drastic 
change in the resulting solar carbon abundance
highlights the importance of having accurate
atomic data for abundance analyses.  
This is especially relevant in the context of the
solar modelling problem, wherein standard models of the solar interior,
adopting the solar chemical composition of 
\citet{2009ARA&amp;A..47..481A}, fail to
reproduce key empirical constraints, including the depth of the convection zone
and interior sound speed that are precisely inferred from helioseismic
observations \citep{2008PhR...457..217B,2019ApJ...881..103Z}.
Extra opacity in the
solar interior near the boundary of the convection zone would resolve the
problem \citep{2015Natur.517...56B}.
Carbon contributes about $5\%$ of the opacity in this region
\citep{2012ApJ...745...10B}, so a higher carbon abundance
would help alleviate the problem, albeit only very slightly.

\section{Conclusions}\label{sec:conclusion}
In the present work, 
energy levels and transition data of E1 transitions are computed for \citoiv{} using the MCDHF and RCI methods. 
Special attention is paid to the computation of transition data involving high Rydberg states by employing an alternative orbital optimization approach.

The accuracy of the predicted excitation energies is evaluated by comparing
with experimental data provided by the NIST database. The average relative differences of the computed
energy levels compared with the NIST data are 0.41\%, 0.081\%, 0.041\%, and 0.0044\%, respectively, for \citoiv{}.
The accuracy of the transition data is evaluated based on the relative differences
of the computed transition rates in the length and velocity gauges, which is given by the quantity $dT$, and by extensive comparisons with previous theoretical and experimental results.
For most of the strong transitions in \citoiv{}, the $dT$ values are less than 5\%. The mean $dT$ for all presented E1 transitions are 8.05\% ($\sigma$ = 0.12), 7.20\% ($\sigma$ = 0.13), 1.77\% ($\sigma$ = 0.050), and 0.28\% ($\sigma$ = 0.0059), respectively, for \citoiv{}. 
Particularly, for strong transitions with $A >$ $10^6$ s$^{-1}$, the mean $dT$
is 1.68\% ($\sigma$ = 0.020), 1.53\% ($\sigma$ = 0.023), 0.297\% ($\sigma$ = 0.010), and 0.205\% ($\sigma$ = 0.0041), respectively, for \citoiv{}. 
By employing alternative optimization schemes of the radial orbitals, the uncertainties $dT$ of the computed transition data for transitions involving high Rydberg states are significantly reduced.
The agreement between computed transition properties, e.g., line strengths, transition rates, and lifetimes, and experimental values is overall good. 
The exception is the weak transitions, e.g., the IC transitions, for which the strong cancellation effects are important; however, these effects cannot be properly considered in the present calculations.
The present calculations are extended to high Rydberg
states that are not covered by previous accurate
calculations and this is of special importance in various astrophysical applications.

The accurate and extensive sets of
atomic data for \citoiv{} are publicly available for
use by the astronomy community.
These data should be useful for opacity calculations
and for models of stellar structures and interiors.
They should also be useful to non-LTE spectroscopic analyses of
both early-type and late-type stars.

\section*{Acknowledgements}
This work is supported by the Swedish research
council under contracts 2015-04842, 2016-04185, 2016-03765, and 2020-03940, and by the Knut and Alice Wallenberg Foundation
under the project grant KAW 2013.0052.
Some of the computations were enabled by resources provided by 
the Swedish National Infrastructure for Computing (SNIC) at 
the Multidisciplinary Center for Advanced Computational Science (UPPMAX) 
and at the High Performance Computing Center North (HPC2N) partially 
funded by the Swedish Research Council through grant agreement no. 
2018-05973. 
This work was also supported by computational resources provided by 
the Australian Government through the 
National Computational Infrastructure (NCI) under the 
National Computational Merit Allocation Scheme (NCMAS),
under project y89.
We thank Nicolas Grevesse for insightful comments on an earlier
version of this manuscript. 
We would also like to thank the anonymous referee for her/his useful comments that helped
improve the original manuscript.
\section*{Data Availability}

The full tables of energy levels (Table \ref{tab:CI-IV_energy}) and transition data (Tables \ref{tab:CI_tr} -- \ref{tab:CIV_tr}) are available in the online Supplementary Material.
 



\bibliographystyle{mnras}
\bibliography{refs} 




\appendix

\section{Additional tables}

\begin{table*}
\caption{\label{tab:CI-IV_energy} Wave function composition (up to three $LS$ components with a contribution $>$ 0.02 of the total wave function) in $LS$-coupling, energy levels (in cm$^{-1}$), and lifetimes (in s; given in length ($\tau_l$) and velocity ($\tau_v$) gauges) for \citoiv{}. Energy levels are given relative to the ground state and compared with NIST data \citep{NIST_ASD}. The full table is available online.}  
\begin{tabular}{llllllllllll}
\hline\midrule
Species & No. & State & $LS$-composition & $E_{RCI} $ & $E_{NIST} $ & $\tau_l$ & $\tau_v$  \\ 
\midrule
C I & 1   & $\mathrm{2s^{2}2p^{2}(^{3}_{2}P)~^{3}P_{0}}$              & 0.88 + 0.03~$\mathrm{2s^{2}2p~^{2}P\,7p~^{3}P}$                                                   & 0            &                                        0        &             &              \\
C I & 2   & $\mathrm{2s^{2}2p^{2}(^{3}_{2}P)~^{3}P_{1}}$              & 0.88 + 0.03~$\mathrm{2s^{2}2p~^{2}P\,7p~^{3}P}$                                                   & 16           &                                        16    	&             &              \\
C I & 3   & $\mathrm{2s^{2}2p^{2}(^{3}_{2}P)~^{3}P_{2}}$              & 0.88 + 0.03~$\mathrm{2s^{2}2p~^{2}P\,7p~^{3}P}$                                                   & 43           &                                        43    	&             &              \\
C I & 4   & $\mathrm{2s^{2}2p^{2}(^{1}_{2}D)~^{1}D_{2}}$              & 0.85 + 0.05~$\mathrm{2s^{2}2p~^{2}P\,7p~^{1}D}$ + 0.03~$\mathrm{2s^{2}2p~^{2}P\,3p~^{1}D}$               & 10~275       &                                        10~193 	&             &              \\
C I & 5   & $\mathrm{2s^{2}2p^{2}(^{1}_{0}S)~^{1}S_{0}}$              & 0.78 + 0.06~$\mathrm{2s^{2}2p~^{2}P\,7p~^{1}S}$ + 0.06~$\mathrm{2p^{4}(^{1}_{0}S)~^{1}S}$                  & 21~775       &                                        21~648 	&             &              \\
C~I & 6   & $\mathrm{2s~^{2}S\,2p^{3}(^{4}_{3}S)~^{5}S_{2}^{\circ}}$    & 0.93 + 0.04~$\mathrm{2s~^{2}S\,2p^{2}(^{3}_{2}P)~^{4}P\,7p~^{5}S^{\circ}}$                          & 33~859       &                                        33~735      &   3.00E-02  &  1.26E-02\\
C~I & 7   & $\mathrm{2s^{2}2p~^{2}P\,3s~^{3}P_{0}^{\circ}}$           & 0.91 + 0.04~$\mathrm{2p^{3}(^{2}_{1}P)~^{2}P\,3s~^{3}P^{\circ}}$                                    & 60~114       &                                        60~333      &   3.00E-09  &  3.04E-09\\
C~I & 8   & $\mathrm{2s^{2}2p~^{2}P\,3s~^{3}P_{1}^{\circ}}$           & 0.91 + 0.04~$\mathrm{2p^{3}(^{2}_{1}P)~^{2}P\,3s~^{3}P^{\circ}}$                                    & 60~133       &                                        60~353      &   3.00E-09  &  3.04E-09\\
C~I & 9   & $\mathrm{2s^{2}2p~^{2}P\,3s~^{3}P_{2}^{\circ}}$           & 0.91 + 0.04~$\mathrm{2p^{3}(^{2}_{1}P)~^{2}P\,3s~^{3}P^{\circ}}$                                    & 60~174       &                                        60~393      &   3.00E-09  &  3.04E-09\\
C~I & 10  & $\mathrm{2s^{2}2p~^{2}P\,3s~^{1}P_{1}^{\circ}}$           & 0.92 + 0.04~$\mathrm{2p^{3}(^{2}_{1}P)~^{2}P\,3s~^{1}P^{\circ}}$                                    & 61~750       &                                        61~982      &   2.78E-09  &  2.83E-09\\
-- & --& --& -- & -- & -- & --& -- \\
\bottomrule
\end{tabular}
\end{table*}

\input{tr_data}

\captionsetup[longtable]{labelfont=bf,labelsep=period}

\begin{multicols}{2}
\end{multicols}
\onecolumn
\begin{longtable}{llccccccccccccccc}
\caption{\label{tab:com_exp} Comparison of relative line strengths ($S$), weighted oscillator strengths ($gf$), and lifetimes ($\tau$), or transition probabilities ($A$), with other theoretical work and experimental results for \citoiv{}. The present values from the MCDHF/RCI calculations are given in the Babushkin(length) gauge. The values in the parentheses are the relative differences between the length and velocity gauges. The references for the experiments are shown in the last column. Note that the sums of the line strengths $S$ have been normalized to 100 for each multiplet in \ion{C}{I}.}\\ 
\hline
\midrule
&&\ion{C}{I} && & &\\
\midrule
Transition array     & Mult.  & $J_u-J_l$ & \multicolumn{4}{c}{$S$ (a.u. of a$_0^2$e$^2$)}  \\
\cmidrule{4-7}
& & & MCDHF/RCI & CIV3$^{(a)}$ & Expt.& Expt. \\
\midrule
$\mathrm{ 2s^22p3p - 2s^22p3s }$ & $\mathrm{^3D -~^3P^o }$ &3 - 2 &  46.67(1.3\%)   &    46.72    &    46.3 $\pm$ 1.8     $^{(b)}$ &       \\ 
       	       	      	             & &2 - 1   &  25.29(1.2\%)   &    25.43    &      25.5 $\pm$ 1.2   $^{(b)}$ &       \\ 
       	       	      	             & &1 - 0   &  11.25(1.2\%)   &    11.29    &      11.8 $\pm$ 0.5   $^{(b)}$ &       \\ 
       	       	      	             & &2 - 2   &  8.047(1.3\%)   &    7.898    &      7.67 $\pm$ 0.38  $^{(b)}$ &       \\
\bottomrule
\end{longtable}
{$^{(a)}$\cite{Hibbert1993}; $^{(b)}$\cite{MUSIELOK1997395}; $^{(c)}$\cite{Bacawski_2001}; $^{(d)}$\cite{Golly_2003}; $^{(e)}$\cite{Tachiev_2000}; $^{(f)}$\cite{Reistad_1986}; $^{(g)}$\cite{Trabert_1999}; $^{(h)}$\cite{Tachiev_1999}; $^{(j)}$\cite{FISCHER1998119}; $^{(k)}$\cite{KNYSTAUTAS197175}; $^{(l)}$\cite{Donnelly1978}; $^{(m)}$\cite{Buchet1973}; $^{(n)}$\cite{Jacques1980}; $^{(o)}$\cite{Peach1988}.}
\begin{multicols}{2}
\end{multicols}

\begin{ThreePartTable}
\scriptsize
\begin{longtable}{@{}llccccccccccccccc}
\caption{\label{tab:CI_com} Comparison of line strengths ($S$) and transition rates ($A$) with other theoretical results for \ion{C}{I}. The present values from the MCDHF/RCI calculations are given in the Babushkin(length) gauge. The wavenumber $\Delta E$ and wavelength $\lambda$
values are taken from the NIST database. The
estimated uncertainties $dT$ of the transition rates are given as percentages in parentheses.}\\ 
\hline
\midrule
Transition array     & Mult.  & $J_u-J_l$ & $\Delta E$ & $\lambda$ & \multicolumn{2}{c}{MCDHF/RCI} && \multicolumn{2}{c}{Spline FCS$^{(a)}$} && \multicolumn{2}{c}{MCHF-BP$^{(b)}$} \\
\cmidrule{6-7}\cmidrule{9-10}\cmidrule{12-13}
& & & ($\mathrm{cm^{-1}}$) & (\AA)& $S$ (a.u. of a$_0^2$e$^2$) & $A$ (s$^{-1}$)& & $S$ (a.u. of a$_0^2$e$^2$) &$A$ (s$^{-1}$)&&$S$ (a.u. of a$_0^2$e$^2$)&$A$ (s$^{-1}$) \\

\midrule
\endfirsthead
\caption{Continued.}\\ 
\hline\midrule
Transition array     & Mult.  & $J_u-J_l$ & $\Delta E$ & $\lambda$ & \multicolumn{2}{c}{MCDHF/RCI} && \multicolumn{2}{c}{Spline FCS$^{(a)}$} && \multicolumn{2}{c}{MCHF-BP$^{(b)}$} \\
\cmidrule{6-7}\cmidrule{9-10}\cmidrule{12-13}
& & & ($\mathrm{cm^{-1}}$) & (\AA)& $S$ (a.u. of a$_0^2$e$^2$) & $A$ (s$^{-1}$)& & $S$ (a.u. of a$_0^2$e$^2$) &$A$ (s$^{-1}$)&&$S$ (a.u. of a$_0^2$e$^2$)&$A$ (s$^{-1}$) \\

\midrule
\endhead
\midrule
\endfoot
$\mathrm{ 2s^22p3d - 2s^22p^2} $ & $\mathrm{^3F^o -~^3P }$&3 - 2 & 78172 & 1279.228 & 	5.79E-02   &   7.92E+06(0.1\%)    &      &   4.52E-02   &   6.24E+06   &      &   8.20E-02   &   1.14E+07   \\   
			              & &2 - 2 & 78155 & 1279.498 & 	1.27E-02   &   2.43E+06(0.3\%)    &      &   9.62E-03   &   1.86E+06   &      &   1.16E-02   &   2.25E+06   \\   
			              & &2 - 1 & 78182 & 1279.056 & 	1.08E-02   &   2.07E+06(0.1\%)    &      &   8.95E-03   &   1.73E+06   &      &   2.08E-02   &   4.04E+06   \\   
$\mathrm{ 2s^22p4d - 2s^22p^2 }$ & $\mathrm{^3F^o -~^3P }$&3 - 2 & 83717 & 1194.488 & 	4.48E-02   &   7.52E+06(0.1\%)    &      &   3.99E-02   &   6.77E+06   &      &      &                      \\   
			              & &2 - 2 & 83709 & 1194.614 & 	1.59E-03   &   3.74E+05(0.5\%)    &      &   1.44E-03   &   3.42E+05   &      &      &                      \\   
			              & &2 - 1 & 83736 & 1194.229 & 	1.97E-02   &   4.64E+06(0.2\%)    &      &   1.73E-02   &   4.10E+06   &      &      &                      \\   
$\mathrm{ 2s^22p5d - 2s^22p^2 }$ & $\mathrm{^3F^o -~^3P }$&3 - 2 & 86283 & 1158.966 & 	4.17E-02   &   7.67E+06(0.3\%)    &      &   3.88E-02   &   7.21E+06   &      &      &                      \\   
			              & &2 - 2 & 86274 & 1159.094 & 	1.38E-03   &   3.55E+05(0.3\%)    &      &   1.36E-03   &   3.54E+05   &      &      &                      \\
\bottomrule
\end{longtable}
{$^{(a)}$\cite{Zatsarinny_2002}; $^{(b)}$\cite{Fischer_2006}.}
\end{ThreePartTable}

\begin{ThreePartTable}
\scriptsize
\begin{longtable}{@{}llccccccccccccccc}
\caption{\label{tab:CII_com} Comparison of line strengths ($S$) and transition rates ($A$) with other theoretical results for \ion{C}{II}. The present values from the MCDHF/RCI calculations are given in the Babushkin(length) gauge. The wavenumber $\Delta E$ and wavelength $\lambda$
values are taken from the NIST database. The
estimated uncertainties $dT$ of the transition rates are given as percentages in parentheses.}\\ 
\hline
\midrule
Transition array     & Mult.  & $J_u-J_l$ & $\Delta E$ & $\lambda$ & \multicolumn{2}{c}{MCDHF/RCI} && \multicolumn{2}{c}{MCHF-BP$^{(a)}$}  && CIV3$^{(b)}$\\
\cmidrule{6-7}\cmidrule{9-10}
& & & ($\mathrm{cm^{-1}}$) & (\AA)& $S$ (a.u. of a$_0^2$e$^2$) & $A$ (s$^{-1}$)& & $S$ (a.u. of a$_0^2$e$^2$) &$A$ (s$^{-1}$) && $A$ (s$^{-1}$)\\

\midrule
\endfirsthead
\caption{Continued.}\\ 
\hline\midrule
Transition array     & Mult.  & $J_u-J_l$ & $\Delta E$ & $\lambda$ & \multicolumn{2}{c}{MCDHF/RCI} && \multicolumn{2}{c}{MCHF-BP$^{(a)}$}  && CIV3$^{(b)}$\\
\cmidrule{6-7}\cmidrule{9-10}
& & & ($\mathrm{cm^{-1}}$) & (\AA)& $S$ (a.u. of a$_0^2$e$^2$) & $A$ (s$^{-1}$)& & $S$ (a.u. of a$_0^2$e$^2$) &$A$ (s$^{-1}$) && $A$ (s$^{-1}$)\\

\midrule
\endhead
\midrule
\endfoot
$ \mathrm{2s2p^2 - 2s^22p }$ & $\mathrm{^2D -~^2P^o }$&5/2 - 3/2 & 74866 & 1335.708 & 	2.03E+00   &   2.89E+08(0.2\%)      &&   2.03E+00   &   2.90E+08   &&   2.89E+08  \\
			          & &3/2 - 3/2 & 74869 & 1335.663 & 	2.24E-01   &   4.79E+07($<$0.05\%)      &&   2.24E-01   &   4.80E+07   &&   4.79E+07  \\
			          & &3/2 - 1/2 & 74932 & 1334.532 & 	1.13E+00   &   2.42E+08(0.1\%)      &&   1.13E+00   &   2.43E+08   &&   2.42E+08  \\
$ \mathrm{2s2p^2 - 2s^22p }$ & $\mathrm{^2S -~^2P^o }$&1/2 - 3/2 & 96430 & 1037.018 & 	1.62E+00   &   1.48E+09(0.7\%)      &&   1.61E+00   &   1.47E+09   &&   1.47E+09  \\
			          & &1/2 - 1/2 & 96493 & 1036.337 & 	8.18E-01   &   7.48E+08(0.6\%)      &&   8.11E-01   &   7.43E+08   &&   7.46E+08  \\
$ \mathrm{2s2p^2 - 2s^22p }$ & $\mathrm{^2P -~^2P^o }$&1/2 - 3/2 & 110560 & 904.480 & 	9.96E-01   &   1.37E+09(0.3\%)      &&   9.96E-01   &   1.37E+09   &&   1.36E+09  \\
			          & &3/2 - 3/2 & 110602 & 904.142 & 	4.95E+00   &   3.41E+09(0.4\%)      &&   4.94E+00   &   3.41E+09   &&   3.38E+09  \\
			          & &1/2 - 1/2 & 110624 & 903.962 & 	1.97E+00   &   2.72E+09(0.4\%)      &&   1.97E+00   &   2.72E+09   &&   2.69E+09  \\
\bottomrule
\end{longtable}
{$^{(a)}$\cite{Tachiev_2000}; $^{(b)}$\cite{CORREGE200419}.}
\end{ThreePartTable}

\begin{ThreePartTable}
\scriptsize{
\begin{longtable}{@{}llccccccccccccccc}
\caption{\label{tab:CIII_com} Comparison of line strengths ($S$) and transition rates ($A$) with other theoretical results for \ion{C}{III}. The present values from the MCDHF/RCI calculations are given in the Babushkin(length) gauge. The wavenumber $\Delta E$ and wavelength $\lambda$
values are taken from the NIST database. The
estimated uncertainties $dT$ of the transition rates are given as percentages in parentheses.}\\
\hline
\midrule
Transition array     & Mult.  & $J_u-J_l$ & $\Delta E$ & $\lambda$ & \multicolumn{2}{c}{MCDHF/RCI} && \multicolumn{2}{c}{MCHF-BP$^{(a)}$} && \multicolumn{2}{c}{\textsc{Grasp}$^{(b)}$} \\
\cmidrule{6-7}\cmidrule{9-10}\cmidrule{12-13}
& & & ($\mathrm{cm^{-1}}$) & (\AA)& $S$ (a.u. of a$_0^2$e$^2$) & $A$ (s$^{-1}$)& & $S$ (a.u. of a$_0^2$e$^2$) &$A$ (s$^{-1}$) & & $S$ (a.u. of a$_0^2$e$^2$) &$A$ (s$^{-1}$) \\

\midrule
\endfirsthead
\caption{Continued.}\\ 
\hline\midrule
Transition array     & Mult.  & $J_u-J_l$ & $\Delta E$ & $\lambda$ & \multicolumn{2}{c}{MCDHF/RCI} && \multicolumn{2}{c}{MCHF-BP$^{(a)}$} && \multicolumn{2}{c}{\textsc{Grasp}$^{(b)}$} \\
\cmidrule{6-7}\cmidrule{9-10}\cmidrule{12-13}
& & & ($\mathrm{cm^{-1}}$) & (\AA)& $S$ (a.u. of a$_0^2$e$^2$) & $A$ (s$^{-1}$)& & $S$ (a.u. of a$_0^2$e$^2$) &$A$ (s$^{-1}$) & & $S$ (a.u. of a$_0^2$e$^2$) &$A$ (s$^{-1}$) \\

\midrule
\endhead
\midrule
\endfoot
$ \mathrm{2s2p - 2s^2 }$ & $\mathrm{^ 1P^o - ~^ 1S }$&1 - 0 & 102352 & 977.020 & 	2.44E+00   &   1.77E+09(0.1\%)    &      &   2.44E+00   &   1.77E+09   &      &   2.38E+00   &   2.15E+09 \\
$ \mathrm{2s3p - 2s^2 }$ & $\mathrm{^ 1P^o - ~^ 1S }$&1 - 0 & 258931 & 386.203 & 	3.06E-01   &   3.59E+09($<$0.05\%)    &      &   3.06E-01   &   3.59E+09   &      &   2.70E-01   &   3.16E+09 \\
$ \mathrm{2s3p - 2s^2 }$ & $\mathrm{^ 3P^o - ~^ 1S }$&1 - 0 & 259711 & 385.043 & 	4.29E-05   &   5.08E+05(0.5\%)    &      &   4.37E-05   &   5.18E+05   &      &   1.17E-02   &   1.36E+08 \\
$ \mathrm{2p^2 - 2s2p }$ & $\mathrm{^ 3P - ~^ 3P^o }$&1 - 2 & 85007 & 1176.370 & 	1.32E+00   &   5.48E+08($<$0.05\%)    &      &   1.32E+00   &   5.50E+08   &      &   1.33E+00   &   5.95E+08 \\
			         & &0 - 1 & 85034 & 1175.987 & 	1.05E+00   &   1.32E+09(0.1\%)    &      &   1.05E+00   &   1.32E+09   &      &   1.07E+00   &   1.43E+09 \\
			         & &2 - 2 & 85054 & 1175.711 & 	3.95E+00   &   9.89E+08(0.1\%)    &      &   3.95E+00   &   9.91E+08   &      &   4.00E+00   &   1.07E+09 \\
			         & &1 - 1 & 85063 & 1175.590 & 	7.90E-01   &   3.30E+08(0.1\%)    &      &   7.89E-01   &   3.31E+08   &      &   8.00E-01   &   3.58E+08 \\
			         & &1 - 0 & 85087 & 1175.263 & 	1.05E+00   &   4.40E+08(0.1\%)    &      &   1.05E+00   &   4.41E+08   &      &   1.07E+00   &   4.78E+08 \\
\bottomrule
\end{longtable}
{$^{(a)}$\cite{Tachiev_1999}; $^{(b)}$\cite{Aggarwal_2015}.}
}
\end{ThreePartTable}

\begin{ThreePartTable}
\begin{longtable}{@{}llccccccccccccccc}
\caption{\label{tab:CIV_com} Comparison of line strengths ($S$) and transition rates ($A$) with other theoretical results for \ion{C}{IV}. The present values from the MCDHF/RCI calculations are given in the Babushkin(length) gauge. The wavenumber $\Delta E$ and wavelength $\lambda$
values are taken from the NIST database. The
estimated uncertainties $dT$ of the transition rates are given as percentages in parentheses.}\\
\hline
\midrule
Transition array     & Mult.  & $J_u-J_l$ & $\Delta E$ & $\lambda$ & \multicolumn{2}{c}{MCDHF/RCI} && \multicolumn{2}{c}{MCHF-BP$^{(a)}$} \\
\cmidrule{6-7}\cmidrule{9-10}
& & & ($\mathrm{cm^{-1}}$) & (\AA)& $S$ (a.u. of a$_0^2$e$^2$) & $A$ (s$^{-1}$)& & $S$ (a.u. of a$_0^2$e$^2$) &$A$ (s$^{-1}$) \\

\midrule
\endfirsthead
\caption{Continued.}\\ 
\hline\midrule
Transition array     & Mult.  & $J_u-J_l$ & $\Delta E$ & $\lambda$ & \multicolumn{2}{c}{MCDHF/RCI} && \multicolumn{2}{c}{MCHF-BP$^{(a)}$} \\
\cmidrule{6-7}\cmidrule{9-10}
& & & ($\mathrm{cm^{-1}}$) & (\AA)& $S$ (a.u. of a$_0^2$e$^2$) & $A$ (s$^{-1}$)& & $S$ (a.u. of a$_0^2$e$^2$) &$A$ (s$^{-1}$) \\

\midrule
\endhead
\midrule
\endfoot
$ \mathrm{2p - 2s} $ & $\mathrm{^ 2P^o - ~^ 2S }$&1/2 - 1/2 & 64484 & 1550.772 & 	9.68E-01   &   2.63E+08($<$0.05\%)  &      &   9.68E-01   &   2.63E+08 \\
			     & &3/2 - 1/2 & 64591 & 1548.187 & 	1.94E+00   &   2.65E+08($<$0.05\%)  &      &   1.94E+00   &   2.65E+08 \\
$\mathrm{ 3p - 2s }$ & $\mathrm{^ 2P^o - ~^ 2S }$&1/2 - 1/2 & 320050 & 312.451 & 	1.39E-01   &   4.63E+09($<$0.05\%)  &      &   1.40E-01   &   4.64E+09 \\
			     & &3/2 - 1/2 & 320081 & 312.420 & 	2.78E-01   &   4.62E+09($<$0.05\%)  &      &   2.79E-01   &   4.63E+09 \\
$\mathrm{ 3s - 2p }$ & $\mathrm{^ 2S - ~^ 2P^o }$&1/2 - 3/2 & 238257 & 419.714 & 	2.07E-01   &   2.84E+09($<$0.05\%)  &      &   2.07E-01   &   2.84E+09 \\
			     & &1/2 - 1/2 & 238365 & 419.525 & 	1.03E-01   &   1.42E+09($<$0.05\%)  &      &   1.03E-01   &   1.42E+09 \\
$\mathrm{ 3d - 2p }$ & $\mathrm{^ 2D - ~^ 2P^o }$&3/2 - 3/2 & 260288 & 384.190 & 	3.26E-01   &   2.91E+09($<$0.05\%)  &      &   3.26E-01   &   2.91E+09 \\
			     & &5/2 - 3/2 & 260298 & 384.174 & 	2.94E+00   &   1.75E+10($<$0.05\%)  &      &   2.94E+00   &   1.75E+10 \\
\bottomrule
\end{longtable}
\begin{tablenotes}
\item {$^{(a)}$\cite{FISCHER1998119}.}
\end{tablenotes}
\end{ThreePartTable}


\bsp	
\label{lastpage}
\end{document}

%% file: scheme.tex
\begin{table*}
\caption{\label{tab:MR} Summary of the computational schemes for \citoiv{}. The first column displays the configurations {of the targeted states}. MR {and AS, respectively,} denote the multireference {sets and the active sets of orbitals} used {in} the MCDHF and RCI calculations, {and} ${N_\mathrm{CSFs}}$ are the numbers of generated {CSFs} {in} the {final} RCI calculations, for the even (e) and {the} odd (o) parity states.}
\begin{tabular}{@{}llcllllll}
\hline
\midrule
{Targeted} configurations     & MR   &     AS     &   $N_\mathrm{{CSFs}}$ \\
\midrule
&\multicolumn{2}{l}{\ion{C}{I}, $\mathrm{N_{levels}}=100$}  & & & \\ 
\midrule
$\mathrm{2s2p^3}$                                 &   $\mathrm{2s2p^3}$                                & \{11s,10p,10d,9f,  & e: 14 941 842   \\
$\mathrm{2s^22p}\{n_1\mathrm{s},n_2\mathrm{p},n_3\mathrm{d,4f}\}$                 & $\mathrm{2s^22p}\{n_1\mathrm{s},n_2\mathrm{p},n_3\mathrm{d,4f}\}$                 & 7g,6h\}  & o: 15 572 953   \\
($3\le n_1\le6$, $2\le n_2\le5$, $3\le n_3\le5$) &  ($3\le n_1\le6$, $2\le n_2\le6$, $3\le n_3\le5$)&    &     \\
                                              &  $\mathrm{2p^3}\{n_1\mathrm{s},n_2\mathrm{p},n_3\mathrm{d}\}$ &   &   \\
                                              &  ($3\le n_1\le6$, $3\le n_2\le5$, $3\le n_3\le6$) &   &    \\
                                              &  $\mathrm{2s2p^2\{3s,3p,4p,6p,6d,7s\}}$ & &   \\
                                              &  $\mathrm{2s2p}\{n_1\mathrm{s},n_2\mathrm{p},n_3\mathrm{d,4f\}6d}$  & &  \\
                                              &  ($3\le n_1\le6$, $3\le n_2\le5$, $3\le n_3\le5$) &   &    \\
\midrule
&\multicolumn{2}{l}{\ion{C}{II}, $\mathrm{N_{levels}}=69$}   & \\ 
\midrule
$\mathrm{2s^2}nl$($n\le6,l\le4$)     & $\mathrm{2s2p^2}$, $\mathrm{2s^2}\{n_1\mathrm{s},n_2\mathrm{p},n_3\mathrm{d},n_4\mathrm{f},n_5\mathrm{g}\}$             &  \{$\mathrm{{14s,14p,12d,12f}},$ &  e: {6~415~798} \\  
$\mathrm{2s^27l}$($l\le3$)			& $(3 \le n_1 \le 9, 2 \le n_2 \le 9, 3 \le n_3 \le 7$,     & $\mathrm{10g,8h}$\}            &  o: {4~988~973}\\
$\mathrm{2s2p^2}$, $\mathrm{2p^3}$,	        & $4 \le n_4 \le 7, 5 \le n_5 \le 6)$                      &   & \\
$\mathrm{2s2p3s}$, $\mathrm{2s2p3p}$          & $\mathrm{2p^3}$, $\mathrm{2p^2}\{n_1\mathrm{s},n_2\mathrm{p},n_3\mathrm{d},n_4\mathrm{f},n_5\mathrm{g}\}$               & & \\
                            & $(3 \le n_1 \le 9, 4 \le n_2 \le 9, 3 \le n_3 \le 7$,     & & \\
                            & $4 \le n_4 \le 7, 5 \le n_5 \le 6)$                      & & \\
                            & {$\mathrm{2s2p3s}$, $\mathrm{2s2p3p}$}                     & & \\
\midrule
&\multicolumn{2}{l}{\ion{C}{III}, $\mathrm{N_{levels}}=114$}   & \\ 
\midrule
$\mathrm{2s}nl$($n\le7, l\le4$)         &   $\mathrm{2s}nl$ ($n\le7, l\le4$)        &   \{$\mathrm{12s,12p,12d,12f,}$ & e: 1 578 620 \\
$\mathrm{2p^2}$, $\mathrm{2p\{3s,3p,3d\}}$    &   $\mathrm{2p^2}$, $\mathrm{2p\{3s,3p,3d\}}$    &      $\mathrm{11g,8h}$\}  & o: 1 274 147 \\
\midrule
&\multicolumn{2}{l}{\ion{C}{IV}, $\mathrm{N_{levels}}=53$}  & \\ 
\midrule
$\mathrm{1s^2}nl$ ($n\le8, l\le4$)	& $\mathrm{1s^2}nl$ ($n\le8, l\le4$)	 &  \{$\mathrm{14s,14p,14d,12f,12g,}$  &      e: 1 077 872 \\
$\mathrm{1s^26h}$			& $\mathrm{1s^26h}$			&    $\mathrm{8h,7i}$\} &o: 1 287 706 \\
\bottomrule
\end{tabular}
\end{table*}

%% file: tr_data.tex
\begin{table*}
\caption{\label{tab:CI_tr} Electric dipole transition data for \ion{C}{I} from present calculations. Upper and lower states, wavenumber, $\Delta E$, wavelength, $\lambda$, line strength, $S$, weighted oscillator strength, $gf$, transition probability, $A$, together with the relative difference between two gauges of $A$ values, $dT$, provided by the present MCDHF/RCI calculations are shown in the table. Wavelength and wavenumber values are from the NIST database \citep{NIST_ASD} when available. Wavelengths and wavenumbers marked with * are from the present calculations. Only the first ten rows are shown; the full table is available online.}
\centering
\begin{tabular}{llccccccccccc}
\hline\midrule
Upper & Lower &  $\Delta E$(cm$^{-1}$)  & $\lambda$ (\AA) & $S$ (a.u. of a$_0^2$e$^2$) & $gf$ & $A$ (s$^{-1}$) & $dT$ \\ 
\midrule
$\mathrm{2s^22p5d~^3D_{  2  }^o}$    & $\mathrm{2s^22p^2~^3P_{  1  }}$   &  86373 &      1157.769 &    1.025E-01  &   2.679E-02 &   2.647E+07  &      0.004\\
$\mathrm{2s^22p5d~^3D_{  1  }^o}$    & $\mathrm{2s^22p^2~^3P_{  0  }}$   &  86362 &      1157.909 &    8.568E-02  &   2.240E-02 &   3.689E+07  &      0.003\\
$\mathrm{2s^22p5d~^3D_{  3  }^o}$    & $\mathrm{2s^22p^2~^3P_{  2  }}$   &  86354 &      1158.018 &    3.197E-01  &   8.358E-02 &   5.897E+07  &      0.003\\
$\mathrm{2s^22p6s~^3P_{  2  }^o}$    & $\mathrm{2s^22p^2~^3P_{  1  }}$   &  86352 &      1158.038 &    1.273E-01  &   3.328E-02 &   3.287E+07  &      0.002\\
$\mathrm{2s^22p5d~^3D_{  1  }^o}$    & $\mathrm{2s^22p^2~^3P_{  1  }}$   &  86346 &      1158.130 &    4.687E-02  &   1.225E-02 &   2.017E+07  &      0.002\\
$\mathrm{2s^22p5d~^3D_{  2  }^o}$ & $\mathrm{2s^22p^2~^3P_{  2  }}$    &    86346 &       1158.131 &   1.049E-01 &   2.742E-02 &   2.708E+07 &  0.000\\
$\mathrm{2s^22p6s~^3P_{  1  }^o}$ & $\mathrm{2s^22p^2~^3P_{  0  }}$    &    86331 &       1158.324 &   2.442E-02 &   6.380E-03 &   1.050E+07 &  0.001\\
$\mathrm{2s^22p6s~^3P_{  2  }^o}$ & $\mathrm{2s^22p^2~^3P_{  2  }}$    &    86325 &       1158.400 &   2.619E-02 &   6.844E-03 &   6.756E+06 &  0.001\\
$\mathrm{2s^22p5d~^3D_{  1  }^o}$ & $\mathrm{2s^22p^2~^3P_{  2  }}$    &    86319 &       1158.492 &   1.367E-03 &   3.571E-04 &   5.875E+05 &  0.002\\
$\mathrm{2s^22p6s~^3P_{  1  }^o}$ & $\mathrm{2s^22p^2~^3P_{  1  }}$    &    86315 &       1158.544 &   6.630E-03 &   1.732E-03 &   2.849E+06 &  0.004\\
-- & --& --& -- & -- & -- & --& -- \\
\bottomrule
\end{tabular}
\end{table*}

\begin{table*}
\caption{\label{tab:CII_tr} Electric dipole transition data for \ion{C}{II} from present calculations. Upper and lower states, wavenumber, $\Delta E$, wavelength, $\lambda$, line strength, $S$, weighted oscillator strength, $gf$, transition probability, $A$, together with the relative difference between two gauges of $A$ values, $dT$, provided by the present MCDHF/RCI calculations are shown in the table. Wavelength and wavenumber values are from the NIST database \citep{NIST_ASD} when available. Only the first ten rows are shown; the full table is available online.}
\centering
\begin{tabular}{llccccccccccc}
\hline\midrule
Upper & Lower &  $\Delta E$(cm$^{-1}$)  & $\lambda$ (\AA) & $S$ (a.u. of a$_0^2$e$^2$) & $gf$ & $A$ (s$^{-1}$) & $dT$ \\ 
\midrule
$\mathrm{2s2p3p~^2D_{ 3/2  }}$   & $\mathrm{2s^22p~^2P_{ 1/2  }^o}$ &  188581 &    530.275 &   8.159E-02 &   4.673E-02 &   2.771E+08 &         0.015\\
$\mathrm{2s2p3p~^2D_{ 5/2  }}$   & $\mathrm{2s^22p~^2P_{ 3/2  }^o}$ &  188551 &    530.359 &   1.515E-01 &   8.678E-02 &   3.430E+08 &         0.015\\
$\mathrm{2s2p3p~^2D_{ 3/2  }}$   & $\mathrm{2s^22p~^2P_{ 3/2  }^o}$ &  188517 &    530.454 &   1.661E-02 &   9.511E-03 &   5.636E+07 &         0.015\\
$\mathrm{2s^27d~^2D_{ 3/2  }}$   & $\mathrm{2s^22p~^2P_{ 1/2  }^o}$ &  187353 &    533.752 &   1.094E-01 &   6.223E-02 &   3.637E+08 &         0.007\\
$\mathrm{2s^27d~^2D_{ 5/2  }}$   & $\mathrm{2s^22p~^2P_{ 3/2  }^o}$ &  187289 &    533.933 &   1.943E-01 &   1.104E-01 &   4.300E+08 &         0.007\\
$\mathrm{2s^27d~^2D_{ 3/2  }}$   & $\mathrm{2s^22p~^2P_{ 3/2  }^o}$  & 187289 &      533.933 &   2.205E-02 &   1.254E-02 &   7.321E+07 &    0.007\\
$\mathrm{2s2p3p~^4P_{ 3/2  }}$   & $\mathrm{2s^22p~^2P_{ 1/2  }^o}$  & 186443 &      536.355 &   1.779E-07 &   1.007E-07 &   5.830E+02 &    0.017\\
$\mathrm{2s2p3p~^4P_{ 1/2  }}$   & $\mathrm{2s^22p~^2P_{ 1/2  }^o}$  & 186427 &      536.402 &   6.007E-07 &   3.400E-07 &   3.936E+03 &    0.039\\
$\mathrm{2s2p3p~^4P_{ 5/2  }}$   & $\mathrm{2s^22p~^2P_{ 3/2  }^o}$  & 186402 &      536.473 &   1.227E-05 &   6.942E-06 &   2.678E+04 &    0.020\\
$\mathrm{2s2p3p~^4P_{ 3/2  }}$   & $\mathrm{2s^22p~^2P_{ 3/2  }^o}$  & 186380 &      536.537 &   1.327E-06 &   7.506E-07 &   4.343E+03 &    0.027\\
-- & --& --& -- & -- & -- & --& -- \\
\bottomrule
\end{tabular}
\end{table*}

\begin{table*}
\caption{\label{tab:CIII_tr} Electric dipole transition data for \ion{C}{III} from present calculations. Upper and lower states, wavenumber, $\Delta E$, wavelength, $\lambda$, line strength, $S$, weighted oscillator strength, $gf$, transition probability, $A$, together with the relative difference between two gauges of $A$ values, $dT$, provided by the present MCDHF/RCI calculations are shown in the table. Wavelength and wavenumber values are from the NIST database \citep{NIST_ASD} when available. Wavelengths and wavenumbers marked with * are from the present calculations. Only the first ten rows are shown; the full table is available online.}  
\centering
\begin{tabular}{llccccccccccc}
\hline\midrule
Upper & Lower &  $\Delta {E}$(cm$^{-1}$)  &
$\lambda$ (\AA) & $S$ (a.u. of a$_0^2$e$^2$) & $gf$ & $A$ (s$^{-1}$) &
$dT$ \\  
\midrule
$\mathrm{2s7p~^3P_{  1  }^o}$ & $\mathrm{2s^2~^1S_{  0  }}$  &  365034*&    273.947*&      7.670E-07    &  8.505E-07 &  2.520E+04 	&  0.005\\                    
$\mathrm{2s7p~^1P_{  1  }^o}$ & $\mathrm{2s^2~^1S_{  0  }}$  &  364896 &    274.051 &      1.043E-02    &  1.156E-02 &  3.423E+08  &  0.010\\
$\mathrm{2s6p~^1P_{  1  }^o}$ & $\mathrm{2s^2~^1S_{  0  }}$  &  357109 &    280.026 &      1.593E-02    &  1.728E-02 &  4.901E+08  &  0.001\\
$\mathrm{2s6p~^3P_{  1  }^o}$ & $\mathrm{2s^2~^1S_{  0  }}$  &  357050 &    280.073 &      5.265E-06    &  5.711E-06 &  1.619E+05  &  0.004\\
$\mathrm{2p3d~^1P_{  1  }^o}$ & $\mathrm{2s^2~^1S_{  0  }}$  &  346712 &    288.423 &      9.317E-04    &  9.818E-04 &  2.627E+07  &  0.003\\
$\mathrm{2s5p~^3P_{  1  }^o}$ & $\mathrm{2s^2~^1S_{  0  }}$   &  344236 &       290.498 &   3.900E-07 &   4.079E-07 &   1.075E+04 &   0.018\\
$\mathrm{2s5p~^1P_{  1  }^o}$ & $\mathrm{2s^2~^1S_{  0  }}$   &  343258 &       291.326 &   4.551E-02 &   4.747E-02 &   1.244E+09 &   0.000\\
$\mathrm{2p3d~^3P_{  1  }^o}$ & $\mathrm{2s^2~^1S_{  0  }}$   &  340127 &       294.007 &   7.645E-07 &   7.903E-07 &   2.035E+04 &   0.005\\
$\mathrm{2p3d~^3D_{  1  }^o}$ & $\mathrm{2s^2~^1S_{  0  }}$   &  337655 &       296.159 &   7.267E-07 &   7.458E-07 &   1.893E+04 &   0.005\\
$\mathrm{2s4p~^1P_{  1  }^o}$ & $\mathrm{2s^2~^1S_{  0  }}$   &  322404 &       310.170 &   3.480E-02 &   3.409E-02 &   7.884E+08 &   0.000\\
-- & --& --& -- & -- & -- & --& -- \\
\bottomrule
\end{tabular}
\end{table*}

\begin{table*}
\caption{\label{tab:CIV_tr} Electric dipole transition data for \ion{C}{IV} from present calculations. Upper and lower states, wavenumber, $\Delta E$, wavelength, $\lambda$, line strength, $S$, weighted oscillator strength, $gf$, transition probability, $A$, together with the relative difference between two gauges of $A$ values, $dT$, provided by the present MCDHF/RCI calculations are shown in the table. Wavelength and wavenumber values are from the NIST database \citep{NIST_ASD}. Only the first ten rows are shown; the full table is available online.}
\centering
\begin{tabular}{llccccccccccc}
\hline\midrule
Upper & Lower &  $\Delta E$(cm$^{-1}$)  & $\lambda$ (\AA) & $S$ (a.u. of a$_0^2$e$^2$) & $gf$ & $A$ (s$^{-1}$) & $dT$ \\ 
\midrule
$\mathrm{8p~^2P_{ 3/2  }^o}$ & $\mathrm{2s~^2S_{ 1/2  }}$   &      492479 &             203.054 &  5.029E-03   &   7.523E-03 &   3.043E+08 &        0.004\\
$\mathrm{8p~^2P_{ 1/2  }^o}$ & $\mathrm{2s~^2S_{ 1/2  }}$   &      492477 &             203.055 &  2.517E-03   &   3.766E-03 &   3.046E+08 &        0.004\\
$\mathrm{7p~^2P_{ 3/2  }^o}$ & $\mathrm{2s~^2S_{ 1/2  }}$   &      483950 &             206.633 &  7.885E-03   &   1.159E-02 &   4.527E+08 &        0.001\\
$\mathrm{7p~^2P_{ 1/2  }^o}$ & $\mathrm{2s~^2S_{ 1/2  }}$   &      483948 &             206.634 &  3.946E-03   &   5.801E-03 &   4.532E+08 &        0.001\\
$\mathrm{6p~^2P_{ 3/2  }^o}$ & $\mathrm{2s~^2S_{ 1/2  }}$   &      470778 &             212.414 &  1.349E-02   &   1.930E-02 &   7.132E+08 &        0.000\\
$\mathrm{6p~^2P_{ 1/2  }^o}$ & $\mathrm{2s~^2S_{ 1/2  }}$   & 470775 &        212.416 &   6.753E-03 &   9.657E-03 &   7.139E+08 &  0.000\\
$\mathrm{5p~^2P_{ 3/2  }^o}$ & $\mathrm{2s~^2S_{ 1/2  }}$   & 448862 &        222.785 &   2.644E-02 &   3.604E-02 &   1.211E+09 &  0.000\\
$\mathrm{5p~^2P_{ 1/2  }^o}$ & $\mathrm{2s~^2S_{ 1/2  }}$   & 448855 &        222.789 &   1.323E-02 &   1.804E-02 &   1.212E+09 &  0.000\\
$\mathrm{8d~^2D_{ 3/2  }}$   & $\mathrm{2p~^2P_{ 1/2  }^o}$ & 428244 &        233.511 &   1.264E-02 &   1.644E-02 &   5.027E+08 &  0.002\\
$\mathrm{8d~^2D_{ 5/2  }}$   & $\mathrm{2p~^2P_{ 3/2  }^o}$ & 428136 &        233.570 &   2.275E-02 &   2.958E-02 &   6.028E+08 &  0.002\\

-- & --& --& -- & -- & -- & --& -- \\
\bottomrule
\end{tabular}
\end{table*}

%% file: main.bbl
\begin{thebibliography}{}
\makeatletter
\relax
\def\mn@urlcharsother{\let\do\@makeother \do\$\do\&\do\#\do\^\do\_\do\%\do\~}
\def\mn@doi{\begingroup\mn@urlcharsother \@ifnextchar [ {\mn@doi@}
  {\mn@doi@[]}}
\def\mn@doi@[#1]#2{\def\@tempa{#1}\ifx\@tempa\@empty \href
  {http://dx.doi.org/#2} {doi:#2}\else \href {http://dx.doi.org/#2} {#1}\fi
  \endgroup}
\def\mn@eprint#1#2{\mn@eprint@#1:#2::\@nil}
\def\mn@eprint@arXiv#1{\href {http://arxiv.org/abs/#1} {{\tt arXiv:#1}}}
\def\mn@eprint@dblp#1{\href {http://dblp.uni-trier.de/rec/bibtex/#1.xml}
  {dblp:#1}}
\def\mn@eprint@#1:#2:#3:#4\@nil{\def\@tempa {#1}\def\@tempb {#2}\def\@tempc
  {#3}\ifx \@tempc \@empty \let \@tempc \@tempb \let \@tempb \@tempa \fi \ifx
  \@tempb \@empty \def\@tempb {arXiv}\fi \@ifundefined
  {mn@eprint@\@tempb}{\@tempb:\@tempc}{\expandafter \expandafter \csname
  mn@eprint@\@tempb\endcsname \expandafter{\@tempc}}}

\bibitem[\protect\citeauthoryear{Aggarwal \& Keenan}{Aggarwal \&
  Keenan}{2015}]{Aggarwal_2015}
Aggarwal K.~M.,  Keenan F.~P.,  2015, \mn@doi [Monthly Notices of the Royal
  Astronomical Society] {10.1093/mnras/stv684}, 450, 1151

\bibitem[\protect\citeauthoryear{{Alexeeva}, {Sadakane}, {Nishimura}, {Du}  \&
  {Hu}}{{Alexeeva} et~al.}{2019}]{2019ApJ...884..150A}
{Alexeeva} S.,  {Sadakane} K.,  {Nishimura} M.,  {Du} J.,   {Hu} S.,  2019,
  \mn@doi [\apj] {10.3847/1538-4357/ab41fa}, \href
  {https://ui.adsabs.harvard.edu/abs/2019ApJ...884..150A} {884, 150}

\bibitem[\protect\citeauthoryear{{Amarsi}, {Barklem}, {Collet}, {Grevesse}  \&
  {Asplund}}{{Amarsi} et~al.}{2019}]{2019A&A...624A.111A}
{Amarsi} A.~M.,  {Barklem} P.~S.,  {Collet} R.,  {Grevesse} N.,   {Asplund} M.,
   2019, \mn@doi [\aap] {10.1051/0004-6361/201833603}, \href
  {http://adsabs.harvard.edu/abs/2019A%26A...624A.111A} {624, A111}

\bibitem[\protect\citeauthoryear{{Asplund}, {Grevesse}, {Sauval}  \&
  {Scott}}{{Asplund} et~al.}{2009}]{2009ARA&amp;A..47..481A}
{Asplund} M.,  {Grevesse} N.,  {Sauval} A.~J.,   {Scott} P.,  2009, \mn@doi
  [\araa] {10.1146/annurev.astro.46.060407.145222}, \href
  {http://adsabs.harvard.edu/abs/2009ARA%26A..47..481A} {47, 481}

\bibitem[\protect\citeauthoryear{Bacawski, Wujec  \& Musielok}{Bacawski
  et~al.}{2001}]{Bacawski_2001}
Bacawski A.,  Wujec T.,   Musielok J.,  2001, \mn@doi [Physica Scripta]
  {10.1238/physica.regular.064a00314}, 64, 314

\bibitem[\protect\citeauthoryear{{Bailey} et~al.,}{{Bailey}
  et~al.}{2015}]{2015Natur.517...56B}
{Bailey} J.~E.,  et~al., 2015, \mn@doi [\nat] {10.1038/nature14048}, \href
  {http://adsabs.harvard.edu/abs/2015Natur.517...56B} {517, 56}

\bibitem[\protect\citeauthoryear{{Basu} \& {Antia}}{{Basu} \&
  {Antia}}{2008}]{2008PhR...457..217B}
{Basu} S.,  {Antia} H.~M.,  2008, \mn@doi [\physrep]
  {10.1016/j.physrep.2007.12.002}, \href
  {http://adsabs.harvard.edu/abs/2008PhR...457..217B} {457, 217}

\bibitem[\protect\citeauthoryear{Berkner, Cooper, Kaplan  \& Pyle}{Berkner
  et~al.}{1965}]{BERKNER196535}
Berkner K.,  Cooper W.,  Kaplan S.,   Pyle R.,  1965, \mn@doi [Physics Letters]
  {https://doi.org/10.1016/0031-9163(65)90390-2}, 16, 35

\bibitem[\protect\citeauthoryear{{Blancard}, {Coss{\'e}}  \&
  {Faussurier}}{{Blancard} et~al.}{2012}]{2012ApJ...745...10B}
{Blancard} C.,  {Coss{\'e}} P.,   {Faussurier} G.,  2012, \mn@doi [\apj]
  {10.1088/0004-637X/745/1/10}, \href
  {https://ui.adsabs.harvard.edu/abs/2012ApJ...745...10B} {745, 10}

\bibitem[\protect\citeauthoryear{Bogdanovich, Karpu{\v{s}}kiene  \&
  Rancova}{Bogdanovich et~al.}{2007}]{Bogdanovich_2007}
Bogdanovich P.,  Karpu{\v{s}}kiene R.,   Rancova O.,  2007, \mn@doi [Physica
  Scripta] {10.1088/0031-8949/75/5/014}, 75, 669

\bibitem[\protect\citeauthoryear{{Boldt}}{{Boldt}}{1963}]{1963ZNatA..18.1107B}
{Boldt} G.,  1963, \mn@doi [Zeitschrift Naturforschung Teil A]
  {10.1515/zna-1963-1009}, \href
  {https://ui.adsabs.harvard.edu/abs/1963ZNatA..18.1107B} {18, 1107}

\bibitem[\protect\citeauthoryear{Buchet-Poulizac \& Buchet}{Buchet-Poulizac \&
  Buchet}{1973a}]{Buchet_Poulizac_1973}
Buchet-Poulizac M.~C.,  Buchet J.~P.,  1973a, \mn@doi [Physica Scripta]
  {10.1088/0031-8949/8/1-2/006}, 8, 40

\bibitem[\protect\citeauthoryear{Buchet-Poulizac \& Buchet}{Buchet-Poulizac \&
  Buchet}{1973b}]{Buchet1973}
Buchet-Poulizac M.~C.,  Buchet J.~P.,  1973b, \mn@doi [Physica Scripta]
  {10.1088/0031-8949/8/1-2/006}, 8, 40

\bibitem[\protect\citeauthoryear{{Caffau}, {Ludwig}, {Bonifacio}, {Faraggiana},
  {Steffen}, {Freytag}, {Kamp}  \& {Ayres}}{{Caffau}
  et~al.}{2010}]{2010A&A...514A..92C}
{Caffau} E.,  {Ludwig} H.~G.,  {Bonifacio} P.,  {Faraggiana} R.,  {Steffen} M.,
   {Freytag} B.,  {Kamp} I.,   {Ayres} T.~R.,  2010, \mn@doi [\aap]
  {10.1051/0004-6361/200912227}, \href
  {https://ui.adsabs.harvard.edu/abs/2010A&A...514A..92C} {514, A92}

\bibitem[\protect\citeauthoryear{{Chen} et~al.,}{{Chen}
  et~al.}{2020}]{2020ApJ...889..157C}
{Chen} X.,  et~al., 2020, \mn@doi [\apj] {10.3847/1538-4357/ab66c7}, \href
  {https://ui.adsabs.harvard.edu/abs/2020ApJ...889..157C} {889, 157}

\bibitem[\protect\citeauthoryear{Corrégé \& Hibbert}{Corrégé \&
  Hibbert}{2004}]{CORREGE200419}
Corrégé G.,  Hibbert A.,  2004, \mn@doi [Atomic Data and Nuclear Data Tables]
  {https://doi.org/10.1016/j.adt.2003.11.002}, 86, 19

\bibitem[\protect\citeauthoryear{{Cunto} \& {Mendoza}}{{Cunto} \&
  {Mendoza}}{1992}]{1992RMxAA..23..107C}
{Cunto} W.,  {Mendoza} C.,  1992, \rmxaa, \href
  {https://ui.adsabs.harvard.edu/abs/1992RMxAA..23..107C} {23, 107}

\bibitem[\protect\citeauthoryear{{Cunto}, {Mendoza}, {Ochsenbein}  \&
  {Zeippen}}{{Cunto} et~al.}{1993}]{1993A&A...275L...5C}
{Cunto} W.,  {Mendoza} C.,  {Ochsenbein} F.,   {Zeippen} C.~J.,  1993, \aap,
  \href {https://ui.adsabs.harvard.edu/abs/1993A&A...275L...5C} {275, L5}

\bibitem[\protect\citeauthoryear{Doerfert, Tr\"abert, Wolf, Schwalm  \&
  Uwira}{Doerfert et~al.}{1997}]{Doerfert1997}
Doerfert J.,  Tr\"abert E.,  Wolf A.,  Schwalm D.,   Uwira O.,  1997, \mn@doi
  [Phys. Rev. Lett.] {10.1103/PhysRevLett.78.4355}, 78, 4355

\bibitem[\protect\citeauthoryear{Donnelly, Kernahan  \& Pinnington}{Donnelly
  et~al.}{1978}]{Donnelly1978}
Donnelly K.~E.,  Kernahan J.~A.,   Pinnington E.~H.,  1978, \mn@doi [J. Opt.
  Soc. Am.] {10.1364/JOSA.68.001000}, 68, 1000

\bibitem[\protect\citeauthoryear{Dyall, Grant, Johnson, Parpia  \&
  Plummer}{Dyall et~al.}{1989}]{EOL}
Dyall K.,  Grant I.,  Johnson C.,  Parpia F.,   Plummer E.,  1989, \mn@doi
  [Comput. Phys. Commun.] {https://doi.org/10.1016/0010-4655(89)90136-7}, 55,
  425

\bibitem[\protect\citeauthoryear{{Ekman}, {Godefroid}  \& {Hartman}}{{Ekman}
  et~al.}{2014}]{Ekman2014}
{Ekman} J.,  {Godefroid} M.,   {Hartman} H.,  2014, \mn@doi [Atoms]
  {10.3390/atoms2020215}, \href
  {https://ui.adsabs.harvard.edu/abs/2014Atoms...2..215E} {2, 215}

\bibitem[\protect\citeauthoryear{Fang, Kwong, Wang  \& Parkinson}{Fang
  et~al.}{1993}]{Fang1993}
Fang Z.,  Kwong V. H.~S.,  Wang J.,   Parkinson W.~H.,  1993, \mn@doi [Phys.
  Rev. A] {10.1103/PhysRevA.48.1114}, 48, 1114

\bibitem[\protect\citeauthoryear{Federman \& Zsargo}{Federman \&
  Zsargo}{2001}]{Federman_2001}
Federman S.~R.,  Zsargo J.,  2001, \mn@doi [The Astrophysical Journal]
  {10.1086/321498}, 555, 1020

\bibitem[\protect\citeauthoryear{Fischer}{Fischer}{1994}]{Fischer_1994}
Fischer C.~F.,  1994, \mn@doi [Physica Scripta] {10.1088/0031-8949/49/3/011},
  49, 323

\bibitem[\protect\citeauthoryear{Fischer}{Fischer}{2000}]{Fischer_2000}
Fischer C.~F.,  2000, \mn@doi [Physica Scripta]
  {10.1238/physica.regular.062a00000}, 62, 458

\bibitem[\protect\citeauthoryear{Fischer}{Fischer}{2006}]{Fischer_2006}
Fischer C.~F.,  2006, \mn@doi [Journal of Physics B: Atomic, Molecular and
  Optical Physics] {10.1088/0953-4075/39/9/005}, 39, 2159

\bibitem[\protect\citeauthoryear{Fischer \& Tachiev}{Fischer \&
  Tachiev}{2004}]{FROESEFISCHER20041}
Fischer C.~F.,  Tachiev G.,  2004, \mn@doi [Atomic Data and Nuclear Data
  Tables] {https://doi.org/10.1016/j.adt.2004.02.001}, 87, 1

\bibitem[\protect\citeauthoryear{Fischer, Saparov, Gaigalas  \&
  Godefroid}{Fischer et~al.}{1998}]{FISCHER1998119}
Fischer C.,  Saparov M.,  Gaigalas G.,   Godefroid M.,  1998, \mn@doi [Atomic
  Data and Nuclear Data Tables] {https://doi.org/10.1006/adnd.1998.0788}, 70,
  119

\bibitem[\protect\citeauthoryear{Fischer, Godefroid, Brage, J{\"o}nsson  \&
  Gaigalas}{Fischer et~al.}{2016}]{Fischer2016}
Fischer C.~F.,  Godefroid M.,  Brage T.,  J{\"o}nsson P.,   Gaigalas G.,  2016,
  J. Phys. B: At. Mol. Opt. Phys, 49, 182004

\bibitem[\protect\citeauthoryear{Fischer, Gaigalas, J{\"o}nsson  \&
  Biero\'n}{Fischer et~al.}{2019}]{Grasp2018}
Fischer C.~F.,  Gaigalas G.,  J{\"o}nsson P.,   Biero\'n J.,  2019, \mn@doi
  [Comput. Phys. Commun.] {https://doi.org/10.1016/j.cpc.2018.10.032}, 237, 184

\bibitem[\protect\citeauthoryear{Fleming, Hibbert  \& Stafford}{Fleming
  et~al.}{1994}]{Fleming_1994}
Fleming J.,  Hibbert A.,   Stafford R.~P.,  1994, \mn@doi [Physica Scripta]
  {10.1088/0031-8949/49/3/010}, 49, 316

\bibitem[\protect\citeauthoryear{{Foster}}{{Foster}}{1962}]{1962PPS....79...94F}
{Foster} E.~W.,  1962, \mn@doi [Proceedings of the Physical Society]
  {10.1088/0370-1328/79/1/314}, \href
  {https://ui.adsabs.harvard.edu/abs/1962PPS....79...94F} {79, 94}

\bibitem[\protect\citeauthoryear{{Franchini} et~al.,}{{Franchini}
  et~al.}{2020}]{2020ApJ...888...55F}
{Franchini} M.,  et~al., 2020, \mn@doi [\apj] {10.3847/1538-4357/ab5dc4}, \href
  {https://ui.adsabs.harvard.edu/abs/2020ApJ...888...55F} {888, 55}

\bibitem[\protect\citeauthoryear{{Froese Fischer}}{{Froese
  Fischer}}{2009}]{Fischer2009}
{Froese Fischer} C.,  2009, \mn@doi [Physica Scripta Volume T]
  {10.1088/0031-8949/2009/T134/014019}, \href
  {https://ui.adsabs.harvard.edu/abs/2009PhST..134a4019F} {134, 014019}

\bibitem[\protect\citeauthoryear{Fuhr}{Fuhr}{2006}]{Fuhr2005}
Fuhr J.~R.,  2006, \mn@doi [Journal of the American Chemical Society]
  {10.1021/ja059868l}, 128, 5585

\bibitem[\protect\citeauthoryear{Gaigalas, Fischer, Rynkun  \&
  J\"onsson}{Gaigalas et~al.}{2017}]{Gaigalas2017}
Gaigalas G.,  Fischer C.~F.,  Rynkun P.,   J\"onsson P.,  2017, \mn@doi [Atoms]
  {10.3390/atoms5010006}, 5

\bibitem[\protect\citeauthoryear{Godefroid, Fischer  \& Jönsson}{Godefroid
  et~al.}{2001}]{Godefroid_2001}
Godefroid M.,  Fischer C.~F.,   Jönsson P.,  2001, \mn@doi [Journal of Physics
  B: Atomic, Molecular and Optical Physics] {10.1088/0953-4075/34/6/308}, 34,
  1079

\bibitem[\protect\citeauthoryear{{Goldbach} \& {Nollez}}{{Goldbach} \&
  {Nollez}}{1987}]{1987A&A...181..203G}
{Goldbach} C.,  {Nollez} G.,  1987, \aap, \href
  {https://ui.adsabs.harvard.edu/abs/1987A&A...181..203G} {181, 203}

\bibitem[\protect\citeauthoryear{{Goldbach}, {Martin}  \& {Nollez}}{{Goldbach}
  et~al.}{1989}]{1989A&A...221..155G}
{Goldbach} C.,  {Martin} M.,   {Nollez} G.,  1989, \aap, \href
  {https://ui.adsabs.harvard.edu/abs/1989A&A...221..155G} {221, 155}

\bibitem[\protect\citeauthoryear{Golly, Jazgara  \& Wujec}{Golly
  et~al.}{2003}]{Golly_2003}
Golly A.,  Jazgara A.,   Wujec T.,  2003, \mn@doi [Physica Scripta]
  {10.1238/physica.regular.067a00485}, 67, 485

\bibitem[\protect\citeauthoryear{{Goly} \& {Weniger}}{{Goly} \&
  {Weniger}}{1982}]{Goly1982}
{Goly} A.,  {Weniger} S.,  1982, \mn@doi [\jqsrt]
  {10.1016/0022-4073(82)90004-8}, \href
  {https://ui.adsabs.harvard.edu/abs/1982JQSRT..28..389G} {28, 389}

\bibitem[\protect\citeauthoryear{Grant}{Grant}{1974}]{gauge}
Grant I.~P.,  1974, J. Phys. B: At. Mol. Opt. Phys, 7, 1458

\bibitem[\protect\citeauthoryear{Grant}{Grant}{2007}]{Grant2007}
Grant I.~P.,  2007, Relativistic Quantum Theory of Atoms and Molecules.
Springer, New York

\bibitem[\protect\citeauthoryear{Haris \& Kramida}{Haris \&
  Kramida}{2017}]{Haris_2017}
Haris K.,  Kramida A.,  2017, \mn@doi [The Astrophysical Journal Supplement
  Series] {10.3847/1538-4365/aa86ab}, 233, 16

\bibitem[\protect\citeauthoryear{Hibbert}{Hibbert}{1974}]{Hibbert_1974}
Hibbert A.,  1974, \mn@doi [Journal of Physics B: Atomic and Molecular Physics]
  {10.1088/0022-3700/7/12/004}, 7, 1417

\bibitem[\protect\citeauthoryear{Hibbert}{Hibbert}{1975}]{HIBBERT1975141}
Hibbert A.,  1975, \mn@doi [Computer Physics Communications]
  {https://doi.org/10.1016/0010-4655(75)90103-4}, 9, 141

\bibitem[\protect\citeauthoryear{{Hibbert}, {Biemont}, {Godefroid}  \&
  {Vaeck}}{{Hibbert} et~al.}{1993}]{Hibbert1993}
{Hibbert} A.,  {Biemont} E.,  {Godefroid} M.,   {Vaeck} N.,  1993, \aaps, \href
  {https://ui.adsabs.harvard.edu/abs/1993A&AS...99..179H} {99, 179}

\bibitem[\protect\citeauthoryear{Jacques, Knystautas, Drouin  \& Berry}{Jacques
  et~al.}{1980}]{Jacques1980}
Jacques C.,  Knystautas E.~J.,  Drouin R.,   Berry H.~G.,  1980, \mn@doi
  [Canadian Journal of Physics] {10.1139/p80-149}, 58, 1093

\bibitem[\protect\citeauthoryear{{Jofr{\'e}}, {Jackson}  \& {Tucci
  Maia}}{{Jofr{\'e}} et~al.}{2020}]{2020A&A...633L...9J}
{Jofr{\'e}} P.,  {Jackson} H.,   {Tucci Maia} M.,  2020, \mn@doi [\aap]
  {10.1051/0004-6361/201937140}, \href
  {https://ui.adsabs.harvard.edu/abs/2020A&A...633L...9J} {633, L9}

\bibitem[\protect\citeauthoryear{Jones \& Wiese}{Jones \&
  Wiese}{1984}]{Jones1984}
Jones D.~W.,  Wiese W.~L.,  1984, \mn@doi [Phys. Rev. A]
  {10.1103/PhysRevA.29.2597}, 29, 2597

\bibitem[\protect\citeauthoryear{J{\"o}nsson, Gaigalas, Biero\'n, Fischer  \&
  Grant}{J{\"o}nsson et~al.}{2013}]{GraspV3}
J{\"o}nsson P.,  Gaigalas G.,  Biero\'n J.,  Fischer C.~F.,   Grant I.,  2013,
  \mn@doi [Comput. Phys. Commun.] {https://doi.org/10.1016/j.cpc.2013.02.016},
  184, 2197

\bibitem[\protect\citeauthoryear{Jönsson, Li, Gaigalas  \& Dong}{Jönsson
  et~al.}{2010}]{JONSSON2010}
Jönsson P.,  Li J.,  Gaigalas G.,   Dong C.,  2010, \mn@doi [Atomic Data and
  Nuclear Data Tables] {https://doi.org/10.1016/j.adt.2009.10.001}, 96, 271

\bibitem[\protect\citeauthoryear{Kingston \& Hibbert}{Kingston \&
  Hibbert}{2000}]{Kingston_2000}
Kingston A.~E.,  Hibbert A.,  2000, \mn@doi [Journal of Physics B: Atomic,
  Molecular and Optical Physics] {10.1088/0953-4075/33/4/307}, 33, 693

\bibitem[\protect\citeauthoryear{Knystautas, Barrette, Neveu  \&
  Drouin}{Knystautas et~al.}{1971}]{KNYSTAUTAS197175}
Knystautas E.,  Barrette L.,  Neveu B.,   Drouin R.,  1971, \mn@doi [Journal of
  Quantitative Spectroscopy and Radiative Transfer]
  {https://doi.org/10.1016/0022-4073(71)90164-6}, 11, 75

\bibitem[\protect\citeauthoryear{Kramida, {Yu.~Ralchenko}, Reader  \& {and NIST
  ASD Team}}{Kramida et~al.}{2019}]{NIST_ASD}
Kramida A.,  {Yu.~Ralchenko} Reader J.,   {and NIST ASD Team} 2019, NIST Atomic
  Spectra Database (ver. 5.7.1), [Online]. Available:
  {\tt{https://physics.nist.gov/asd}} [2020, February 11]. National Institute
  of Standards and Technology, Gaithersburg, MD.

\bibitem[\protect\citeauthoryear{{Kwong}, {Fang}, {Gibbons}, {Parkinson}  \&
  {Smith}}{{Kwong} et~al.}{1993}]{Kwong1993}
{Kwong} V. H.~S.,  {Fang} Z.,  {Gibbons} T.~T.,  {Parkinson} W.~H.,   {Smith}
  P.~L.,  1993, \mn@doi [\apj] {10.1086/172845}, \href
  {https://ui.adsabs.harvard.edu/abs/1993ApJ...411..431K} {411, 431}

\bibitem[\protect\citeauthoryear{Li, Jönsson, Dong  \& Gaigalas}{Li
  et~al.}{2010}]{Li_2010}
Li J.,  Jönsson P.,  Dong C.,   Gaigalas G.,  2010, \mn@doi [Journal of
  Physics B: Atomic, Molecular and Optical Physics]
  {10.1088/0953-4075/43/3/035005}, 43, 035005

\bibitem[\protect\citeauthoryear{{Maecker}}{{Maecker}}{1953}]{1953ZPhy..135...13M}
{Maecker} H.,  1953, \mn@doi [Zeitschrift fur Physik] {10.1007/BF01329774},
  \href {https://ui.adsabs.harvard.edu/abs/1953ZPhy..135...13M} {135, 13}

\bibitem[\protect\citeauthoryear{Mickey}{Mickey}{1970}]{MICKEY197077}
Mickey D.,  1970, \mn@doi [Nuclear Instruments and Methods]
  {https://doi.org/10.1016/0029-554X(70)90653-1}, 90, 77

\bibitem[\protect\citeauthoryear{Miller, Wilkerson, Roig  \& Bengtson}{Miller
  et~al.}{1974}]{Miller1974}
Miller M.~H.,  Wilkerson T.~D.,  Roig R.~A.,   Bengtson R.~D.,  1974, \mn@doi
  [Phys. Rev. A] {10.1103/PhysRevA.9.2312}, 9, 2312

\bibitem[\protect\citeauthoryear{Moore \& Gallagher}{Moore \&
  Gallagher}{1993}]{Moore1993}
Moore C.~E.,  Gallagher J.~W.,  1993, Tables of spectra of hydrogen, carbon,
  nitrogen, and oxygen atoms and ions.
Boca Raton : CRC Press

\bibitem[\protect\citeauthoryear{Musielok, Veres  \& Wiese}{Musielok
  et~al.}{1997}]{MUSIELOK1997395}
Musielok J.,  Veres G.,   Wiese W.,  1997, \mn@doi [Journal of Quantitative
  Spectroscopy and Radiative Transfer]
  {https://doi.org/10.1016/S0022-4073(96)00124-0}, 57, 395

\bibitem[\protect\citeauthoryear{Nandi, Bhattacharya, Kurup  \& Prasad}{Nandi
  et~al.}{1996}]{Nandi_1996}
Nandi T.,  Bhattacharya N.,  Kurup M.~B.,   Prasad K.~G.,  1996, \mn@doi
  [Physica Scripta] {10.1088/0031-8949/54/2/011}, 54, 179

\bibitem[\protect\citeauthoryear{{Nieva} \& {Przybilla}}{{Nieva} \&
  {Przybilla}}{2006}]{2006ApJ...639L..39N}
{Nieva} M.~F.,  {Przybilla} N.,  2006, \mn@doi [\apjl] {10.1086/501124}, \href
  {https://ui.adsabs.harvard.edu/abs/2006ApJ...639L..39N} {639, L39}

\bibitem[\protect\citeauthoryear{{Nieva} \& {Przybilla}}{{Nieva} \&
  {Przybilla}}{2008}]{2008A&A...481..199N}
{Nieva} M.~F.,  {Przybilla} N.,  2008, \mn@doi [\aap]
  {10.1051/0004-6361:20078203}, \href
  {https://ui.adsabs.harvard.edu/abs/2008A&A...481..199N} {481, 199}

\bibitem[\protect\citeauthoryear{{Nieva} \& {Przybilla}}{{Nieva} \&
  {Przybilla}}{2012}]{2012A&A...539A.143N}
{Nieva} M.~F.,  {Przybilla} N.,  2012, \mn@doi [\aap]
  {10.1051/0004-6361/201118158}, \href
  {https://ui.adsabs.harvard.edu/abs/2012A&A...539A.143N} {539, A143}

\bibitem[\protect\citeauthoryear{{Nussbaumer} \& {Storey}}{{Nussbaumer} \&
  {Storey}}{1981}]{Nussbaumer1981}
{Nussbaumer} H.,  {Storey} P.~J.,  1981, \aap, \href
  {https://ui.adsabs.harvard.edu/abs/1981A&A....96...91N} {96, 91}

\bibitem[\protect\citeauthoryear{{Nussbaumer} \& {Storey}}{{Nussbaumer} \&
  {Storey}}{1984}]{Nussbaumer1984}
{Nussbaumer} H.,  {Storey} P.~J.,  1984, \aap, \href
  {https://ui.adsabs.harvard.edu/abs/1984A&A...140..383N} {140, 383}

\bibitem[\protect\citeauthoryear{{Olsen}, {Roos}, {J{\o}rgensen}  \&
  {Jensen}}{{Olsen} et~al.}{1988}]{Olsen_AS}
{Olsen} J.,  {Roos} B.~O.,  {J{\o}rgensen} P.,   {Jensen} H. J.~A.,  1988,
  \mn@doi [\jcp] {10.1063/1.455063}, \href
  {https://ui.adsabs.harvard.edu/abs/1988JChPh..89.2185O} {89, 2185}

\bibitem[\protect\citeauthoryear{Papoulia et~al.,}{Papoulia
  et~al.}{2019}]{Papoulia2019}
Papoulia A.,  et~al., 2019, Atoms, 7

\bibitem[\protect\citeauthoryear{Peach, Saraph  \& Seaton}{Peach
  et~al.}{1988}]{Peach1988}
Peach G.,  Saraph H.~E.,   Seaton M.~J.,  1988, \mn@doi [Journal of Physics B:
  Atomic, Molecular and Optical Physics] {10.1088/0953-4075/21/22/006}, 21,
  3669

\bibitem[\protect\citeauthoryear{Pehlivan~Rhodin, Hartman, Nilsson  \&
  J\"onsson}{Pehlivan~Rhodin et~al.}{2017}]{Pehlivan_MgI}
Pehlivan~Rhodin A.,  Hartman H.,  Nilsson H.,   J\"onsson P.,  2017, \mn@doi
  [A\&A] {10.1051/0004-6361/201629849}, 598, A102

\bibitem[\protect\citeauthoryear{{Przybilla}, {Butler}  \&
  {Kudritzki}}{{Przybilla} et~al.}{2001}]{2001A&A...379..936P}
{Przybilla} N.,  {Butler} K.,   {Kudritzki} R.~P.,  2001, \mn@doi [\aap]
  {10.1051/0004-6361:20011384}, \href
  {https://ui.adsabs.harvard.edu/abs/2001A&A...379..936P} {379, 936}

\bibitem[\protect\citeauthoryear{Reistad \& Martinson}{Reistad \&
  Martinson}{1986}]{Reistad1986}
Reistad N.,  Martinson I.,  1986, \mn@doi [Phys. Rev. A]
  {10.1103/PhysRevA.34.2632}, 34, 2632

\bibitem[\protect\citeauthoryear{Reistad, Hutton, Nilsson, Martinson  \&
  Mannervik}{Reistad et~al.}{1986}]{Reistad_1986}
Reistad N.,  Hutton R.,  Nilsson A.~E.,  Martinson I.,   Mannervik S.,  1986,
  \mn@doi [Physica Scripta] {10.1088/0031-8949/34/2/011}, 34, 151

\bibitem[\protect\citeauthoryear{{Richter}}{{Richter}}{1958}]{1958ZPhy..151..114R}
{Richter} J.,  1958, \mn@doi [Zeitschrift fur Physik] {10.1007/BF01344209},
  \href {https://ui.adsabs.harvard.edu/abs/1958ZPhy..151..114R} {151, 114}

\bibitem[\protect\citeauthoryear{Roberts \& Eckerle}{Roberts \&
  Eckerle}{1967}]{Roberts1967}
Roberts J.~R.,  Eckerle K.~L.,  1967, \mn@doi [Phys. Rev.]
  {10.1103/PhysRev.153.87}, 153, 87

\bibitem[\protect\citeauthoryear{{Stonkut{\.{e}}} et~al.,}{{Stonkut{\.{e}}}
  et~al.}{2020}]{2020AJ....159...90S}
{Stonkut{\.{e}}} E.,  et~al., 2020, \mn@doi [\aj] {10.3847/1538-3881/ab6a19},
  \href {https://ui.adsabs.harvard.edu/abs/2020AJ....159...90S} {159, 90}

\bibitem[\protect\citeauthoryear{{Sturesson}, {J{\"o}nsson}  \& {Froese
  Fischer}}{{Sturesson} et~al.}{2007}]{Sturesson_AS}
{Sturesson} L.,  {J{\"o}nsson} P.,   {Froese Fischer} C.,  2007, \mn@doi
  [CoPhC] {10.1016/j.cpc.2007.05.013}, \href
  {https://ui.adsabs.harvard.edu/abs/2007CoPhC.177..539S} {177, 539}

\bibitem[\protect\citeauthoryear{Tachiev \& Fischer}{Tachiev \&
  Fischer}{1999}]{Tachiev_1999}
Tachiev G.,  Fischer C.~F.,  1999, \mn@doi [Journal of Physics B: Atomic,
  Molecular and Optical Physics] {10.1088/0953-4075/32/24/315}, 32, 5805

\bibitem[\protect\citeauthoryear{Tachiev \& Fischer}{Tachiev \&
  Fischer}{2000}]{Tachiev_2000}
Tachiev G.,  Fischer C.~F.,  2000, \mn@doi [Journal of Physics B: Atomic,
  Molecular and Optical Physics] {10.1088/0953-4075/33/13/304}, 33, 2419

\bibitem[\protect\citeauthoryear{Tachiev \& Fischer}{Tachiev \&
  Fischer}{2001}]{Tachiev2001}
Tachiev G.,  Fischer C.~F.,  2001, \mn@doi [Canadian Journal of Physics]
  {10.1139/p01-059}, 79, 955

\bibitem[\protect\citeauthoryear{Träbert, Gwinner, Knystautas, Tordoir  \&
  Wolf}{Träbert et~al.}{1999}]{Trabert_1999}
Träbert E.,  Gwinner G.,  Knystautas E.~J.,  Tordoir X.,   Wolf A.,  1999,
  \mn@doi [Journal of Physics B: Atomic, Molecular and Optical Physics]
  {10.1088/0953-4075/32/17/103}, 32, L491

\bibitem[\protect\citeauthoryear{{VandenBerg}, {Bergbusch}, {Dotter},
  {Ferguson}, {Michaud}, {Richer}  \& {Proffitt}}{{VandenBerg}
  et~al.}{2012}]{2012ApJ...755...15V}
{VandenBerg} D.~A.,  {Bergbusch} P.~A.,  {Dotter} A.,  {Ferguson} J.~W.,
  {Michaud} G.,  {Richer} J.,   {Proffitt} C.~R.,  2012, \mn@doi [\apj]
  {10.1088/0004-637X/755/1/15}, \href
  {https://ui.adsabs.harvard.edu/abs/2012ApJ...755...15V} {755, 15}

\bibitem[\protect\citeauthoryear{Wiese \& Fuhr}{Wiese \&
  Fuhr}{2007a}]{Wiese2007}
Wiese W.~L.,  Fuhr J.~R.,  2007a, \mn@doi [Journal of Physical and Chemical
  Reference Data] {10.1063/1.2740642}, 36, 1287

\bibitem[\protect\citeauthoryear{Wiese \& Fuhr}{Wiese \& Fuhr}{2007b}]{Erratum}
Wiese W.~L.,  Fuhr J.~R.,  2007b, \mn@doi [Journal of Physical and Chemical
  Reference Data] {10.1063/1.2809438}, 36, 1737

\bibitem[\protect\citeauthoryear{Ynnerman \& Fischer}{Ynnerman \&
  Fischer}{1995}]{Ynnerman1995}
Ynnerman A.,  Fischer C.~F.,  1995, \mn@doi [Phys. Rev. A]
  {10.1103/PhysRevA.51.2020}, 51, 2020

\bibitem[\protect\citeauthoryear{Zatsarinny \& Fischer}{Zatsarinny \&
  Fischer}{2002}]{Zatsarinny_2002}
Zatsarinny O.,  Fischer C.~F.,  2002, \mn@doi [Journal of Physics B: Atomic,
  Molecular and Optical Physics] {10.1088/0953-4075/35/22/309}, 35, 4669

\bibitem[\protect\citeauthoryear{{Zhang}, {Li}  \&
  {Christensen-Dalsgaard}}{{Zhang} et~al.}{2019}]{2019ApJ...881..103Z}
{Zhang} Q.-S.,  {Li} Y.,   {Christensen-Dalsgaard} J.,  2019, \mn@doi [\apj]
  {10.3847/1538-4357/ab2f77}, \href
  {https://ui.adsabs.harvard.edu/abs/2019ApJ...881..103Z} {881, 103}

\makeatother
\end{thebibliography}
